% \headline{\hfill {version 8.5}} 

 %%%%%%%%%%%%%%%%%%%%%%%%%%%%%%%%%%%%%%%%%%%%
 % COMMENT OUT PREVIOUS LINE FOR FINAL DRAFT
 %%%%%%%%%%%%%%%%%%%%%%%%%%%%%%%%%%%%%%%%%%%%

% $Id: tsirelson.tex,v 1.5 2004/09/13 20:49:52 rideout Exp $

% Outlining is at end!

% Time-stamp{2006-Jul-06 01:31:27 17580.41007 at umoja.phy.syr.edu}

%: Page Setup for 8.5 x 11 inch paper

 \message{Assuming 8.5 x 11 inch paper.}
 \message{' ' ' ' ' ' ' '}

\magnification=\magstep1                  % \magstep1=1200
\raggedbottom
\parskip=9pt
\overfullrule=0pt 

%: TeX macros

\def\singlespace{\baselineskip=12pt}      % spacing for stuff like abstract
\def\sesquispace{\baselineskip=16pt}      % spacing for main text
\def\extraspace{\baselineskip=20pt}   % spacing for dense text

%% \def\sesquispace{\baselineskip=20pt}      % draft spacing

%%%%%%%% Macros %%%%%%%%%
%\input amstex.tex

%%%%%%%% Macros Rafael %%%%%%%%%
%\input ../../../papers/include/msmacros
%\input ../../../papers/include/mathmacros

%\input mathmacros.greekbold
%\input msmacros
%\input mathmacros

% (Outline "\\\\..." "%%.." )

% This is `msmacros.tex' Time-stamp{2005-Nov-30 05:29:55 17293.32547 at umoja.phy.syr.edu}
%
% The macros here are specifically for manuscripts, and concern such things as
% title, abstract, section titles, page formatting, etc.

% For Spanish accents and umlauts do `tengeneza-tex-acentos'

%% Currenty using subsection titles as in hamilton-crew paper.  
%% Have also taken title-font and sec headings from there.

% Roster of definitions contained herein
%
% \Nobreak
%
% \linebreak = \lbr
% \pagebreak
%
% \BulletItem
%
% \BeginIndentation (possibly make single space: see developing.macros/indentation.tex)
% \EndIndentation
%
% \QuotationBegins (very provisional)
% \QuotationEnds   (very provisional)
% 
% \REMARK	
% \CONJECTURE	
% \THEOREM	(provisional)
% \COROLLARY	(provisional)
% \LEMMA	(provisional)
%
% \footnote (re-defined)
%
% \titlefont (provisional)
% \headingfont (provisional)
% \subheadfont (provisional)
% 
% \author
% \address
% \furtheraddress
% \email
%
% \AbstractBegins
% \AbstractEnds
%
% \section (provisional)
% \subsection (provisional)
%
% \ReferencesBegin
% \reference = \ref
% \sepref
% \journaldata
% \eprint
% \webref (and \webhome)
% \arxiv (this yields a URL)

%%%%%%%%%%%%%%%%%%%%%%%%%%%%%%%%%%%%%%%%%%%%%%%%%%%%%%%%%%%%%
%% Following is to improve appearance of footnotes      
%
 \let\miguu=\footnote
 \def\footnote#1#2{{$\,$\parindent=9pt\baselineskip=13pt%
 \miguu{#1}{#2\vskip -7truept}}}
%
% Notes
%   NOT breaking the line in the middle of the footnote-macro seems crucial!!??
%   Without the 9pt indent, the footnote symbol sticks out to the left.
%   The vskip at the end adjusts separation between multiple footnotes
%      (even 0pt separates them too much!)
% NB
%   Insert an extra \vskip -7pt (say) if phantom last line of footnote
%   spills over onto next page!
%
%%%%%%%%%%%%%%%%%%%%%%%%%%%%%%%%%%%%%%%%%%%%%%%%%%%%%%%%%%%%%

\def\linebreak{\hfil\break}

\def\pagebreak{\vfil\break}

	% to prevent page break (usually works)

\def\BulletItem #1 {\item{$\bullet$}{#1 }}

\def\AbstractBegins
{
 \singlespace                                        % spacing for abstract
 \bigskip\leftskip=1.5truecm\rightskip=1.5truecm     % begin indentation
 \centerline{\bf Abstract}
 \smallskip
 \noindent	% this doesn't seem to take effect over a blank line
 } 
\def\AbstractEnds
{
 \bigskip\leftskip=0truecm\rightskip=0truecm       %  end indentation
 }

\def\ReferencesBegin
{
 \singlespace					   % single spacing
 \vskip 0.5truein
 \centerline           {\bf References}
 \par\nobreak
 \medskip
 \noindent
 \parindent=2pt
 \parskip=6pt			% earlier was 10 pt and then 4pt
 }
 %
 % \raggedright 
 %   Don't use this!
 %   It got even more overfull hboxes, and it ruined the
 %   right-justification of the references!

%%% DO WE WANT A  \ReferencesEnd  too to restore parskip etc?
%%% Maybe not.  Rather a \GlossaryBegins or whatever is meant to follow
%%% the references.

\def\section #1 {\bigskip\noindent{\headingfont #1 }\par\nobreak\noindent}

\def\subsection #1 {\medskip\noindent{\subheadfont #1 }\par\nobreak\noindent}
% \def\subsection #1 {\medskip\noindent[ {\it #1} ]\par\nobreak\smallskip}
% \def\subsection #1 {\medskip\noindent{\it [ #1 ]}\par\nobreak\smallskip} 

%%% try 2pc here (what's a pc anyhow? perhaps a pico?)
\def\reference{\hangindent=1pc\hangafter=1} % to put before each reference

\def\ref{\reference}

 %
 % To separate individual items within a single reference

\def\journaldata#1#2#3#4{{\it #1\/}\phantom{--}{\bf #2$\,$:} $\!$#3 (#4)}
%% \def\journaldata#1#2#3#4{{\it #1 } {\bf #2$\,$:}     #3 (#4)}
%% \def\journaldata#1#2#3#4{{\it #1\/} \  {\bf #2\phantom{.}:} $\!$#3 (#4)}
%% \def\journaldata#1#2#3#4{{\it #1\/}~{\bf #2 $\!$:} $\!$#3 (#4)}
%% \def\journaldata#1#2#3#4{{\it #1 } {\bf #2:} #3 (#4)}
 %
 % Arguments are {journal name}{volume}{pages}{year}
 % (volume can also be {volume plus issue number})
 % A space after #1 seems to do something

 %
 % \def\eprint#1{$\langle$#1\hbox{$\rangle$}}
 %  Was the hbox around { > } really needed here?
 %
 % The following sort of worked:
 % \def\eprint#1{{$\langle$ #1 \hbox{ $\rangle$ }}}
 %
 % \def\eprint #1 {{$\langle$e-print archive: #1 $\rangle$ }}

\def\arxiv#1{\hbox{\tt http://arXiv.org/abs/#1}}

\def\author#1 {\medskip\centerline{\it #1}\bigskip}

\def\REMARK{\noindent {\csmc Remark \ }}

\def\THEOREM{\noindent {\csmc Theorem \ }}
\def\COROLLARY{\noindent {\csmc Corollary \ }}
\def\LEMMA{\noindent {\csmc Lemma \ }}
\def\PROOF{\noindent {\csmc Proof \ }}
\def\DEFINITION{\noindent {\csmc Definition \ }} 

%%%%%%%%%%%%%%%%%%%%%%%%%%%%%%%%%%%%%%%%
%% All of the following are provisional
%%%%%%%%%%%%%%%%%%%%%%%%%%%%%%%%%%%%%%%%

%% In these fonts, maybe better to scale 2 1 1 rather than 2 1 0
%% magstep1 means 12 pt, magstep1 2 means 14.4 pt 

\font\titlefont=cmb10 scaled\magstep2 

\font\headingfont=cmb10 at 12pt
% \font\headingfont=cmb10 at 13pt
% \font\headingfont=cmb10 scaled\magstep1
%% magstep2 is too big. Is magstep1 big enough? (if not, try 13pt)

\font\subheadfont=cmssi10 scaled\magstep1 % sans serif italic
%\font\subheadfont=cmb10 scaled\magstep0

%% Check following with Valdivia paper, when they are working!

\font\csmc=cmcsc10  % caps and small caps

%% Following is for extended quotations, it narrows text and single-spaces
%% Amalgamate this with \BeginIndentation ?
%% Get these from valdivia paper, when they are working!

% \def\QuotationBegins
% {
%  \singlespace
%  \smallskip\leftskip=1.5truecm\rightskip=1.5truecm
%  \noindent	% this doesn't seem to take effect over a blank line
%  } 
% \def\QuotationEnds
% {
%  \smallskip\leftskip=0truecm\rightskip=0truecm
%  \noindent
%  } 
%  % The initial smallskip is needed to activate the rest, apparently!

%% \endinput   
 % Commented out, because it can cause trouble when the present file is
 % included bodily in some other tex file! 

%% END of the file `msmacros' 

% (Outline "\\\\...." "%> ."  "%% .." "%%% .")

% This is `mathmacros.tex'  Time-stamp{2006-Jun-17 21:55:54 17556.45738 at umoja.phy.syr.edu}
%
% The macros here are mainly for mathematical symbols, while
% the file  msmacros.tex  has most of the manuscript formatting macros.

%> Symbols in "openface" font for the integers, reals, etc.

%% In the following, the \hbox seems to be needed when the symbol is
%% used in math mode (which it almost always is, of course)!

\font\openface=msbm10 at10pt
 %
 % If you use 12pt instead of 10pt then it seems to get over-magnified
 % when you specify an overall magnification=1200

\def\Reals         {{\hbox{\openface R}}}

%% Alternatives, in case ``blackboard font'' is unavailable:
 %
 % \def\Integers{{\bf Z}}
 % %   rob's crude open faced Z doesn't seem to work anymmore, so for now just
 % %   use boldface if bbf is unavailable
 % %   \def\Integers{\hbox{${\rm Z \kern 0.3ex \llap{Z}}$}}
 %
 % \def\Reals{{\rm I\!\rm R}}	% open-face R, for real number line
 %
 % \def\Complexes
 %    {{\rm C}\llap{\vrule height6.3pt width1pt depth-.4pt\phantom t}}

%> Symbols in german ("gothic") font

%% Again \hbox is apparently needed when symbol used in math mode!
%% Using \Buchstabe takes care of this automatically

\font\german=eufm10 at 10pt

\def\Buchstabe#1{{\hbox{\german #1}}}

 % symmetrization symbol

%> Named math operators like `sin' or `det' in roman font

%%% Should we add back the spaces in `im' `ker' and `card' ?
%% `Re' and `Im' remain untested

\def\Re  {\mathop{\rm Re}  \nolimits}    % Real part
    % Imaginry part

    % image
    % cardinality

	 % automorphism group
	 % equivalence class
	 % diffeomorphism group

	 % trace

    % curl

%% BEWARE Things above with final space, as once in `Tr', give bad spacing
% when you put a subscript on them, as in: " \tr_{II} R "
%
% NB putting the initial { \, } in the def of `tr' OUTSIDE the {\rm } messes
% up the centering of equations that begin with it !!!

%% Some previous versions
%
% \def\ker{{\rm ker}\,}		% kernel (not needed since built in)
% \def\im{{\rm im}\,}		% image
% \def\card{{\rm card}\,}	% cardinality
% \def\Aut{{\rm Aut}}		% automorphism group
%
% \def\cls{{\rm cls}}		% equivalence class
% \def\Diff{{\rm Diff}}		% diffeomorphism group

% \def\tr{\mathop {{\rm\,Tr\,}} \nolimits}	 % trace

%> Symbols for implication

\def\implies{\Rightarrow}

%% These two cause trouble if you try to use, eg, "\=" as a bar accent
%% over a letter 
%
% \def\=>{\Rightarrow}
% \def\==>{\Longrightarrow}

%> Symbol for box operator (d'Alembertian)
% (good darkness, but a bit too small) (from Eric Woolgar?)
%  (Don't call it `box' since that already exists in TeX as a builtin!) 

 \def\dal{\displaystyle{{\hbox to 0pt{$\sqcup$\hss}}\sqcap}}

%%% shouldn't we make this be a ``mathop'' like the others?

%> Symbols for "less than or of the order of" and its inverse 
%  taken from David Wiltshire

\def\lto{\mathop
        {\hbox{${\lower3.8pt\hbox{$<$}}\atop{\raise0.2pt\hbox{$\sim$}}$}}}
\def\gto{\mathop
        {\hbox{${\lower3.8pt\hbox{$>$}}\atop{\raise0.2pt\hbox{$\sim$}}$}}}
%
% Alternate versions
%
%   for a still better general method see arvinds  "stacksymbol" in the
%   file { ~/ms/texfiles/developing.macros/arvind.macros }
%
% \def\lto { {\raise1pt\hbox{$<$}} \!\!\!\! {\lower4pt\hbox{$\sim$}} }
% \def\gto { {\raise1pt\hbox{$>$}} \!\!\!\! {\lower4pt\hbox{$\sim$}} }
% \def\lto {{\lower4pt\hbox{$\buildrel<\over\sim$}}}
% \def\gto {{\lower4pt\hbox{$\buildrel>\over\sim$}}}

%> the fractions 1/2 1/3 and 1/4 

\def\half{{1 \over 2}}

%> More symbols 

\def\braces#1{ \{ #1 \} }

\def\bra{<\!}			% These seem more useful than the two
\def\ket{\!>}			% commented out ones just below
				% Could, also just redefine \langle and
				% \rangle to be these
	% \def\ket#1{|#1>}      %
	% \def\bra#1{<#1|}      %

		% symbol for isomorphism
		% symbol for isomorphism

	% symbol used eg in f:A-->B

\def\ideq{\equiv}		% triple equal sign

\def\SetOf#1#2{\left\{ #1  \,|\, #2 \right\} }

		% symbol for set-theoretic difference

		% tensor product symbol

\def\interior #1 {  \buildrel\circ\over  #1}     % seems to work
 % alternate
 % \def\interior #1 {{ \buildrel\circ\over{{#1}} }} % works, not too well

% Lie Derivative symbol
  % notice that ${\rm{\it\$}}$ % fails

% semidirect product

%% These are for basis vectors and covectors, with labels (between
%% parentheses) directly over or under the ``kernel''.
%% The order of arguments is: {kernel} {label} {abstract index}
%% usage example: \dualbasisvector{e}{j}{\mu}

\def\basisvector#1#2#3{
 \lower6pt\hbox{
  ${\buildrel{\displaystyle #1}\over{\scriptscriptstyle(#2)}}$}^#3}

\def\alfa{\alpha}

%> tildes, hats, and other accents 

		% define tilde to always be the ``widetilde'' 
\def\bar{\overline}		% define bar to always be wide bar
\def\hat{\widehat}		% define hat to always be the ``widehat'' 

%> spacing of indices, etc 

%%% Can we get around this problem?
%% Following fucks up use of ampersand!  So comment it out for now
%% and make it available separately in:  mathmacros.tensor.indices

%% \def\&{{\phantom a}}		% for spacing tensor indices properly

%> END of the file `mathmacros'

% This is the file `mathmacros.greekbold.tex'
% to be incorporated into mathmacros ultimately
% Time-stamp{2005-Apr-05 23:22:32 16979.22008 at umoja.phy.syr.edu}

%%%%%%%%%%%%%%%%%%%%%%%%%%%%%%%%%%%%%%%%%%%%%%%%%%%%%%%%%%%%%%%%%%%%%%%%
% See the original file in ./archive/ from which this was adapted
% (Some other things there might be useful too)
%   
% The version David Craig sent has some Greek letters that were originally
% lacking here, see ./developing.macros/mathmacros.greekbold.alt.tex
%%%%%%%%%%%%%%%%%%%%%%%%%%%%%%%%%%%%%%%%%%%%%%%%%%%%%%%%%%%%%%%%%%%%%%%%

% W.T. GRANDY, JR. UNIVERSITY OF WYOMING

%: BOLDFACE GREEK (plus a few other stray symbols)

%NOTE: This was found necessary in the implementation used
%      by the company pcTEX---if your version already provides
%      all the Greek letters and symbols in boldface math
%      italic, then you should erase this subset of instructions
%      before using the package with your version of TEX.

% EXAMPLE: boldface alpha is {\bm\alpha}

% It is necessary to redefine the font family for the
% lower-case Greek letters defined in plain.tex

\font\bmit=cmmib10			%Boldface math italic
\font\expo=cmmib10 at 10 true pt	%superscript version (exponent)

\newfam\boldmath 
\def\bm{\fam\boldmath\bmit}

\textfont8=\bmit 
\scriptfont8=\expo 
\scriptscriptfont8=\expo

  \mathchardef\alpha="710B     
  \mathchardef\beta="710C
  \mathchardef\gamma="710D     
  \mathchardef\delta="710E
  \mathchardef\epsilon="710F   
  \mathchardef\zeta="7110
  \mathchardef\eta="7111       
  \mathchardef\theta="7112
  \mathchardef\iota="7113
  \mathchardef\kappa="7114     
  \mathchardef\lambda="7115
  \mathchardef\mu="7116        
  \mathchardef\nu="7117
  \mathchardef\xi="7118        
  \mathchardef\pi="7119
  \mathchardef\rho="711A       
  \mathchardef\sigma="711B
  \mathchardef\tau="711C       
  \mathchardef\upsilon="711D
  \mathchardef\phi="711E
  \mathchardef\chi="711F
  \mathchardef\psi="7120
  \mathchardef\omega="7121     
  \mathchardef\varepsilon="7122
  \mathchardef\vartheta="7123
  \mathchardef\varpi="7124
  \mathchardef\varrho="7125
  \mathchardef\varsigma="7126
  \mathchardef\varphi="7127

  \mathchardef\imath="717B	% dotless i
  \mathchardef\jmath="717C	% dotless j
  \mathchardef\ell="7160
  \mathchardef\partial"7140

% Greek CAPTIALS are already defined this way in plain.tex

%: SOME USEFUL MATHEMATICAL SPECIALTIES

\def\goesto#1{\quad\lower 1ex\overrightarrow{\ssize\hphantom{M}
    #1 \hphantom{M}}\quad}
               % Provides a long right arrow with #1 underneath.

\def\goesblank{{\;\hbox to 25pt{\rightarrowfill}\;}}
               %long right arrow with nothing underneath

% END of the file `mathmacros.greekbold.tex'

%% Try different fonts

\font\titlefont=cmb10 scaled\magstep2

\font\headingfont=cmb10 at 13pt
% \font\headingfont=cmb10 scaled\magstep1
% \font\headingfont=cmb10 scaled\magstep2

\font\subheadfont=cmssi10 scaled\magstep1

\def\section #1 {\bigskip\noindent{\headingfont #1 }\par\nobreak\noindent}
\def\subsection #1 {\medskip\noindent{\subheadfont #1
}\par\nobreak\noindent}

\def\arxiv#1{\hbox{\tt http://arXiv.org/abs/#1}}

% Make next three be bold:

\def\LEMMA#1{\noindent {\bf Lemma #1\ }}
\def\THEOREM#1{\noindent {\bf Theorem #1\ }}
\def\COROLLARY{\noindent {\bf Corollary \ }}

\def\QED{\hfill\hbox{\csmc QED}}

\def\H{\Buchstabe{H}}

\def\frac#1#2{{#1\over#2}}
\def\bfa{{\bm{a}}}
\def\bfb{{\bm{b}}}

\def\Prob{{\rm Prob}}
\def\hOm{\hat\Omega}

\def\hmu{\hat\mu}

% define \bra and \ket to be \langle, \rangle
\def\bra{\langle}
\def\ket{\rangle}

\def\braket#1#2{\bra#1|#2\ket}

\def\QI#1#2#3{#1(#2 \,;\,#3)}

\def\abc{a \cap b \cap c}
\def\alfbetc{\alfa \cap \beta \cap c}
  % bold math symbols
  % bold math italic

\def\csojm{4.1}
\def\qjmso{4.2 }
\def\lsp{4.3 }
\def\thmA{1}
\def\thmB{2}
\def\lemmaA{3.1}
\def\lemmaB{3.2}

%: preprint number(s)
% (Also print version number if called for above)

\phantom{}
\vskip -1.1 true in
\medskip

\rightline{quant-ph/0605008}

\vskip 0.3 true in
\bigskip
\bigskip

%%%%%%%%%%%%%%%% title page %%%%%%%%%%%%%%%%%%%%%%%%%%%%%%%%%%%%

%: Title and authors

\sesquispace
\centerline {\titlefont A Bell Inequality Analog in Quantum Measure Theory}

\vskip 10.mm
\centerline{
David Craig$^{a}$,
Fay Dowker$^{b}$,
Joe Henson$^{c}$,
}
\centerline{
Seth Major$^{d}$,
David Rideout$^{e}$
and
Rafael D.~Sorkin$^{f}$
}

\vfill

\AbstractBegins
One obtains Bell's inequalities if one posits a hypothetical joint
probability distribution, or {\it measure}, whose marginals yield the
probabilities produced by the spin measurements in question.
The existence of a joint measure is in turn equivalent to a certain
%% RDS: local 
causality condition known as ``screening off''.
We show that if one assumes, more generally, a joint
{\it quantal measure}, or ``decoherence functional'', one obtains instead an
analogous inequality weaker by a factor of $\sqrt{2}$.  The proof of this
``Tsirel'son inequality'' is geometrical and rests on the possibility of
associating a Hilbert space to any strongly positive quantal measure.  These
results lead both to a {\it question}: 
``Does a joint measure follow from
some
quantal analog of `screening off'?'', 
and to the {\it observation} that
non-contextual hidden variables are viable in histories-based quantum
mechanics, even if they are excluded classically.
\AbstractEnds

\vfill

%: footnote type addresses

\hrule width 5.cm
\vskip 5mm

%%%{\small }

\noindent $^a$ Le Moyne College, Syracuse New York 13214, USA, and
Hamilton College, Clinton New York 13323, USA.

\noindent $^b$ Blackett Laboratory, Imperial College, London SW7 2AZ, UK,
and Perimeter Institute, Waterloo, Ontario N2L 2Y5, Canada.

\noindent $^c$ Institute for Theoretical Physics, University of Utrecht,
Minnaert Building, Leuvenlaan 4, 3584 CE
Utrecht, The Netherlands

\noindent $^d$ Department of Physics, Hamilton College, Clinton New York
13323, USA.

\noindent $^e$ Blackett Laboratory, Imperial College, London SW7 2AZ, UK.
% and Department of Physics, Hamilton College, Clinton New York 13323, USA.

\noindent $^f$
Perimeter Institute, Waterloo, Ontario N2L 2Y5, Canada,
and Department of Physics, Syracuse University, Syracuse New York 13244,
USA.

\break

%%%%%%%%%%%%%%%%%%%%%%%%%%%%%%%%%%%%%%%%

%: Turn on spacing for body of paper

\sesquispace

\section{I. Introduction}
Thinking of an experiment designed to test the Bell inequalities, we
might picture 
to ourselves
a source emitting a pair of
silver atoms with correlated spins,
and downstream, two Stern-Gerlach analyzers in spacelike
separated regions, $A$ and $B$.  For each setting of the two analyzers
one would obtain a set of $2\times2=4$ experimental probabilities
(frequencies) corresponding to the four possible combinations of
spin-up-or-down.  By differently orienting one or both of the analyzers,
one could similarly produce further sets of four experimental
probabilities.  A collection of probabilities obtained in this way, we
will refer to as a
{\it system of experimental probabilities}.
The Bell inequality [1] (or more precisely its offspring, the
Clauser-Horne-Shimony-Holt-Bell (CHSHB) inequality [2]
[3]) pertains to such a system of experimental
probabilities in the special case obtained by limiting each analyzer to
only two possible settings (say $a$ and $a'$ for the $A$-analyzer, and
$b$ and $b'$ for the $B$-analyzer).

Via a derivation that we recall below, the CHSHB inequality follows
almost immediately from an assumption which we will express by saying
that the given system of experimental probabilities
{\it admits a joint probability distribution}.
To clarify what this means, notice that, a priori,
one has (with two settings each for the analyzers)
four entirely distinct probability distributions,
each living in its own four-element sample space
$\Omega_{\alfa\beta}=\Omega_\alfa\times\Omega_\beta$,
where
$\alfa$ ranges over the settings $a$ or $a'$ of $A$,
$\beta$ ranges over the settings $b$ or $b'$ of $B$,
and
each space $\Omega_\alfa$, $\Omega_\beta$ is a binary sample space,
corresponding to the two possibilities, spin-up/spin-down.
To say that these probabilities admit a joint distribution means that
one can merge the $\Omega_{\alfa\beta}$ into
a single sample space
$$
  \hat\Omega = \Omega_a \times \Omega_{a'} \times \Omega_b \times
\Omega_{b'}
  \eqno(1)
$$
of $2^4=16$ elements,
and that
one can define on $\hat\Omega$ a (not necessarily unique) probability
distribution from which, for example, the probabilities for
$\Omega_{ab}=\Omega_a\times\Omega_b$ follow on summing over the possible
$a'$ and $b'$ outcomes.\footnote{$^\star$}
{That there are 16 experimental probabilities and 16 joint probabilities
 is merely a coincidence.  The two numbers would differ if we
 generalized to particles of higher spin or considered more than two
 settings per analyzer.}
That is, the separate distributions on the spaces $\Omega_{\alfa\beta}$
can be recovered as {\it marginals} from a single probability
distribution on the joint sample space $\hat\Omega$.

In effect one is assigning a meaning to the
so called ``counterfactual''
question,
``What would I find if I could observe all four spin axes $a$,
$a'$, $b$ and $b'$ at once?''
And --- crucially --- one is assuming that
the distributions for the $\Omega_{\alfa\beta}$
(induced as marginals from the joint distribution on $\hat\Omega$)
are
merely
{\it ``revealed''} but not {\it altered}
by the particular way in which the analyzers are set,
the ``context'' of the observation.
For this reason, the assumption of a joint
probability distribution is often alternatively described as the
assumption of 
non-contextual 
``hidden variables'' [4], 
and the
violation of the CHSHB inequality is then described as an
experimental refutation of such 
hidden variables theories.
It is also described as a refutation of 
``local causality'' because a condition of that type implies 
the existence of non-contextual hidden variables.

Thus far,
however,
the implicit context of our discussion has been entirely classical,
and one may wonder to what extent
the relationships we have
just reviewed carry over
to the quantum case.  It might seem that this
question is ill posed, for the lack of a quantal analog of the notion of
joint
probability distribution.  However, if one views quantum mechanics from a
``histories'' standpoint, then it is natural to regard it as a kind of
generalized 
theory of probability or {\it measure} 
(probability being realized mathematically in terms of the 
concept of measure).
Indeed, one can delineate a hierarchy of such 
generalized 
measure theories
[5] in which classical stochastic theories comprise the first
level of the hierarchy and unitary quantum theories --- suitably interpreted
--- are included in the second level. (See also [6].)

Within this second level, the level of ``quantal measures'' or
``decoherence functionals'' (we use the terms interchangeably), one has
a notion of joint quantal measure, in direct analogy to the notion
of joint probability distribution.
We will see that
just as the assumption that the experimental probabilities admit a
joint {\it classical} measure leads to the CHSHB inequality, so the
assumption that they admit a joint {\it quantal} measure leads, almost as
directly, to an analogous but
weaker constraint known as
the Tsirel'son inequality [7].  The main result of this
paper, then, is that the latter inequality can be understood as a direct
analog of the CHSHB inequality if one adopts a histories formulation of
quantum mechanics.  Such a formulation also leads to a geometrical proof
of the inequality that, we believe, has some independent interest in its
own right.\footnote{$^\dagger$}
{A related result, which identifies ``quantum Bell inequalities''
which are necessary conditions for a set of two-qubit states to be the
reduced states of a mixed state of three qubits, appears in
[8].}

Of course, the connection between the CHSHB inequality and the
existence
of a joint measure is far from the whole story in the classical case,
because the strongest  support
for the latter assumption usually comes from
considerations of causality and/or locality.\footnote{$^\flat$}
 {Not all authors distinguish between these concepts, but we try to do so
 consistently here, {\it cf.\ }[9].  By {\it locality} we mean
 the failure of
 physical influences to ``jump over regions of spacetime'', and by {\it
 causality} (in the sense of relativistic causality) we mean their failure
 to
 act outside the future light cone.  For example, a theory containing
``tachyons''
 might be local without being causal.}
Of particular importance in this connection is
the condition known variously as
``local causality'' [10],
``stochastic Einstein locality'' [11],
or ``classical screening off''.

The condition of ``screening off'' on the classical measure asserts that
events in causally unrelated regions of spacetime
become independent
(are ``screened off'' from one another)
when one conditions on a complete specification
of the history in their mutual causal past.
As shown by Fine [4], 
the derivation
of the CHSHB inequality from screening off 
can be viewed as a two-step
process.  
%
% From this point of view, 
% Fine's derivation
% of the CHSHB inequality [R::Fine:1982] can be viewed as a two-step
% process.  
%
First one goes from screening off to the existence of a joint
probability distribution, and then from the latter to the inequality.
(The converse implications are also valid [4].)
The violation in nature of the CHSHB inequality is thus also a
violation of classical screening off.  Usually this is described as a
``failure of locality'', but because screening off is above all a condition
of
relativistic {\it causality},
it might be more appropriate
to rather characterize violation of the CHSHB inequality as a
``failure of (classical) causality''.

Should quantum mechanics, then, be thought of as nonlocal, acausal, or both
---
or is there a sense in which it is neither if seen from an appropriate
vantage
point?
We would have liked, in the present paper, to provide
such a vantage point
by showing that the
classical
threefold
equivalence
among
screening off,
the existence of a joint probability measure,
and the
CHSHB inequality reproduces itself at a higher level (namely level two) as a
relationship
among
{\it quantal} screening off, the existence of a joint
decoherence functional, and the Tsirel'son inequality (which of course is
{\it
not} violated by quantum mechanics).

The proof that a joint quantal measure implies the Tsirel'son inequality
accomplishes this in part, but we are unable to complete the story
in all generality
because we lack a fully convincing formulation
of quantal screening off.  Nevertheless,
we will suggest a candidate condition that closely resembles its
classical analog, and that is formally valid in relativistic quantum
field theory.  We will be able to
prove that any system of experimental
probabilities that admits a joint decoherence functional also admits a
model which obeys
this
screening off condition; but the converse eludes
us, and so we cannot yet assert that screening off is fully equivalent
to a joint measure in the quantal case.
We {\it will}
show, however, that
a causality assumption inherent in standard unitary quantum theory,
namely the commuting of spacelike separated operators, does imply the
existence of a joint measure.  This provides a kind of converse and
shows in particular how our proof of the Tsirel'son inequality
can be
founded on a recognizable causality condition.

\section{II. Quantum mechanics as quantum measure theory}
We briefly summarize the hierarchy of generalized measure theories
described in more detail in [5] [12]
[13].

In a generalized measure theory, there is a sample space $\Omega$ of
possibilities for the system in question.
Normally these are  to be thought of as ``fine grained histories'',
meaning as complete a description of physical reality as is conceivable
in the theory, {\it e.g.} for $n$-particle mechanics a history would be
a set of $n$ trajectories, and for a scalar field theory, a history
would be a field configuration on spacetime.
Predictions about the system --- the dynamical
content of the theory ---  are to be gleaned,
in some way or another, from
a (generalized) measure $\mu$
on $\Omega$
(strictly, on some suitable class of ``measurable'' subsets of
$\Omega$, but we will gloss over this
technicality
here).

Given $\mu$ (a non-negative real-valued set function),
we can construct the
following series of symmetric set functions:
$$
\eqalign{
        I_1(X) & \equiv \mu(X) \cr
      I_2(X,Y) & \equiv \mu(X\sqcup Y) - \mu(X) - \mu(Y) \cr
  I_3(X, Y, Z) & \equiv \mu(X\sqcup Y \sqcup Z) -
      \mu(X\sqcup Y) - \mu(Y\sqcup Z) - \mu(Z\sqcup X) + \mu(X) + \mu(Y) +
\mu(Z) \cr
}
$$
and so on, where $X$, $Y$, $Z$, {\it etc.\ }are disjoint subsets of
$\Omega$,
as indicated by the symbol `$\sqcup$' for disjoint union.

A measure theory of level $k$ is one which satisfies the sum rule
$I_{k+1}=0$.
It is known that this condition implies that all higher sum rules are
automatically satisfied, {\it viz.\ }$I_{k+n}=0$ for all $n\geq 1$.
A level 1 theory is thus
one in which the measure satisfies the usual
Kolmogorov sum rules of classical probability theory,
classical
Brownian motion
being a good example.
A level 2 theory is one in which the Kolmogorov sum rules
may be violated but $I_3$ is nevertheless zero.  Unitary quantum mechanics
satisfies this condition and is an example of a level 2 theory --- which we
dub therefore ``quantum measure theory'' in general.

The existence of a
normalized
quantum measure on $\Omega$ is equivalent to the
existence of a
{\it decoherence functional}
$D(X;Y)$ of pairs of subsets of $\Omega$
satisfying:\footnote{$^\star$}
{The quantity $D(X;Y)$ is
 interpretable as the quantum interference between two sets of
 histories in the case when they are disjoint.
 Notice
 from (2)
 that $\mu$
 determines only the real part of $D$.
 (The imaginary part of $D$
  influences how smaller systems combine to form bigger ones.
  It also
  may affect the consistency/decoherence conditions one
  wishes to impose.
  These issues will be discussed in greater depth elsewhere.)}

\noindent (i) Hermiticity: $D(X;Y) = D(Y;X)^*$ ,\  $\forall X, Y$;

\noindent (ii) Additivity: $D(X\sqcup Y; Z) = D(X;Z) + D(Y;Z)$ ,\
   $\forall X, Y, Z$ with $X$ and $Y$ disjoint;

\noindent (iii) Positivity: $D(X;X)\ge0$ ,\  $\forall X$;
% \noindent (iii) (Weak) positivity: $D(X;X)\ge0$ ,\  $\forall X$;

\noindent (iv) Normalization: $D(\Omega ;\Omega)=1$  .

\noindent The relationship between the quantal measure and
the decoherence functional is
$$
    \mu(X) = D(X ; X).         \eqno(2)
$$

Unless otherwise stated, we will always assume that $D$ satisfies
in addition to (iii)
the condition of {\it strong positivity},
which states that for any finite collection of
(not necessarily disjoint)
subsets $X_1,X_2,\dots X_n$
of $\Omega$,
the $n \times n$
Hermitian
matrix $M_{ij} =D(X_i; X_j)$ is positive
semidefinite
(it has no negative expectation values).
The decoherence functional of ordinary unitary quantum mechanics, for
example, is strongly positive.
Strong positivity
is a powerful requirement
because it
implies
in general
that there is a Hilbert space associated with the quantum
measure,
which turns out to be the standard Hilbert space in the case of
unitary quantum mechanics [14] [15].
Decoherence functionals which merely satisfy condition (iii) above are
termed ``weakly positive''.

In this paper, we will not
enter into
the general question of how to
interpret the quantum measure.
One set of ideas for doing so
goes by the name of ``consistent histories'' or ``decoherent histories''
and attempts in effect to reduce the quantal measure to a classical one
by the imposition of decoherence conditions
[16]
[17]
[18]
[19].
A different attempt at an interpretation,
based on the notions of ``preclusion'' and correlation,
may be found in [12].
For our purposes in this paper,
it will suffice to assume,
where macroscopic measuring instruments are concerned,
that distinct ``pointer readings'' do not interfere
(they ``decohere''),
and
that their measures can
be interpreted as probabilities in the sense of frequencies.

In the sequel, we adopt a usage that seems particularly suitable for
a histories-based
measure theory.
We use the terminology ``an {\it event} in spacetime region
$A$'' to refer to a subset $a\subseteq\Omega$ such that the criterion which
determines whether or not a history $\gamma$ belongs to $a$ refers only to
the
properties of $\gamma$ within $A$ ({\it e.g.\ }if $\gamma$ is a field then
its
restriction to $A$ is
supposed to be
enough information to determine whether
$\gamma\in a$).\footnote{$^\dagger$}
{The term ``event'' is standard in probability theory for a subset of
$\Omega$.
 A subset of histories defined by some common property is termed a
 ``coarse grained history'' in the standard parlance of
 consistent histories quantum theory.
 In this language,
 an event in $A$ is therefore a
 coarse grained history defined by a coarse graining 
 according to properties 
 local to $A$.}

We will
also
assume
that all of our sample spaces
$\Omega$ are finite, so that integrals may be written as sums.
Among other things, this
lets us avoid the main technical complications in the
definition of conditional
probability.

\section{III. Two inequalities, classical and quantal}

\subsection{Classical case}
We rehearse the proof of the CHSHB inequality at level one in the
hierarchy of generalized measures --- {\it i.e.\ }at the classical level.
The experimental context will be that described in the Introduction.
In formalising it, however,
one faces a choice. Namely, one
must
decide
whether or not to include random variables
corresponding to the instrument settings in the analysis.
If one excludes such variables, then one need deal only with the
sample spaces described in the Introduction: the four spaces
$\Omega_{\alfa\beta}$,
the
four
spaces
$\Omega_\alfa$ and $\Omega_\beta$, and the joint sample space
$\hat\Omega$.
(Recall that in our notation,
$\alfa=a$ or $a'$, and
$\beta=b$ or $b'$,
the instrument settings in regions $A$ and $B$.)
On the other hand, if one includes the instrument settings
as variables,
then one
necessarily deals with a
larger
sample space $\Omega$.
We
have chosen to follow
the second approach (which arguably is ``more fully intrinsic'',
in keeping with the philosophy of generalized quantum
mechanics as a theory of closed systems),
and consequently our
discussion
will attribute probabilities not only to the possible outcomes with a
given experimental arrangement, but also to the possible experimental
arrangements themselves.
Nevertheless, the following may
also
be read consistently
as if the first,
more ``minimalist''
approach had been adopted, since, {\it mutatis mutandis}, the
proofs take the same form in both cases.
(One who feels uncomfortable attributing quantitative probabilities to
instrument settings may thus refrain from doing so.)
The
essential
difference
between the two approaches
is that
in the ``minimalist'' reading,
conditional probabilities
like $\Prob(outcome|setting)$ must be understood as primitive objects; they
cannot be resolved into ratios of conditional probabilities like
$\Prob(outcome \cap setting)/\Prob(setting)$.

Consider a sample space, $\Omega$, of histories defined on a
``substratum''
possessing a
background
causal structure
(a spacetime, for example, or a causal set).
Let $A$ and $B$ be two spacelike
separated regions of the
substratum
and denote their causal pasts
by $J^-(A)$ and $J^-(B)$ (where $J^-(A)$ contains $A$ itself).
We are interested in the usual
EPRB setup in which there is a range of possible choices
(to be made ``essentially freely'') of {\it settings} of
some experimental apparatus in $A$ and similarly in $B$.
For example, this range might be the possible directions of the
magnetic field in a Stern-Gerlach apparatus for spin measurements.
For each setting in $A$, the outcome of the
measurement is either $+1$ or $-1$. In the standard example
this would be the measured value of the spin (multiplied by $2/\hbar$)
in the set direction.

Let $M_A$ denote the set of possible settings of the
experimental apparatus in $A$. 
(As mentioned in the Introduction, we limit ourselves to two
 settings, $M_A = \braces{a,a'}$.) 
Each element of $M_A$ is,
in our technical sense,
an event
in $A$
(or more generally in
$J^-(A)\cap J^-(B)^c$,
where
the superscript $c$
denotes complementation),
namely,
that subset of
$\Omega$ containing
those histories
in which the corresponding
experimental setting is made.  The elements of $M_A$
are disjoint.
For each 
element, $\alfa \in M_A$, let
$\alfa_{\pm 1}$ denote the possible outcomes of the measurement
with setting
$\alfa$,
so that
$\alfa=\alfa_{+1}\sqcup \alfa_{-1}$
where $\alfa_{+1}$ ($\alfa_{-1}$) is the set of histories
in which outcome $+1$ ($-1$) obtains.
Similarly, $M_B$ is the set of 
two 
possible
$B$-measurements, 
$\braces{b,b'}$;
each element of $M_B$ is an
event in $B$ (or $J^-(B){\cap}J^-(A)^c$);
and
for each $\beta \in M_B$,
$\beta_{\pm 1}$ are the possible outcomes of the measurement, with
$\beta= \beta_{+1} \sqcup \beta_{-1}$.

The ``law of motion'' of the underlying stochastic process
is assumed to be given by
a classical probability measure $\mu$ on $\Omega$,
and
an expression
like $\mu(\alfa_i\cap\beta_j | \alfa\cap\beta)$
will denote the probability of outcomes $\alfa_i$ and $\beta_j$,
{\it conditional on} the settings being $\alfa$ and $\beta$.
Since we are imagining all our sample spaces as finite, a conditional
probability $\mu(x|y)=\mu(x{\cap}y)/\mu(y)$ is only really meaningful
when $\mu(y)>0$.  In the contrary case, one might define it to be zero,
since $\mu(y)=0\implies\mu(x{\cap}y)=0$ (albeit not when $\mu$ is
quantal!), but for present purposes, it will prove more convenient to
adopt
the convention that $\mu(x|y)$ is simply {\it undefined} when
$\mu(y)$ vanishes.

Let $\hat\Omega$ be
the sixteen-element sample space
(1)
labelled by the 
(16 possible values of the) 
quadruple of binary variables
$(a_i,a'_{i'},b_j,b'_{j'})$,
each of which takes values $\pm1$.
(For brevity we will write
$(a_i,a'_{i'},b_j,b'_{j'}) \equiv (i \, i' \, j \, j')$
where there is no risk of confusion.)
In the Introduction, we called the sixteen numbers
$\mu(a_i\cap b_j|a\cap b)$,
$\mu(a'_{i'}\cap b_j|a' \cap b)$,
$\mu(a_{i}\cap b'_{j'}|a \cap b')$,
$\mu(a'_{i'}\cap b'_{j'}|a' \cap b')$,
a ``system of experimental probabilities'', and we agreed to say that
these numbers {\it admit a joint probability distribution}
if and only if
there exists a
classical measure, $\hat\mu$ on $\hat\Omega$, such that
$$
  \mu(a_i \cap b_j | a\cap b )= \sum_{i'j'} \hat\mu(i \, i' \, j \, j') \ ,
  \eqno(3)
$$
and similarly for every
other
$(\alfa,\beta)$ pair.
It is now easy to prove the CHSHB inequality.

\THEOREM{\thmA}
Let $\Omega$ and $\mu$ be as described above
and
assume that the resulting system of experimental probabilities admits a
joint probability distribution $\hat\mu$ on $\hat\Omega$ satisfying
the condition (3) on its marginals.
Define the correlation functions
$$
  X(\alfa,\beta) \equiv \sum_{ij} i \cdot j \cdot
  \mu(\alfa_i \cap {\beta_j} \,|\, \alfa \cap \beta)
  \eqno{(4)}
$$
for 
$\alfa = a,a'$, $\beta=b,b'$.
Then
$$
   | \; X(a,b) + X(a', b) + X(a, b') - X(a',b') \; | \le 2  \ .
   \eqno(5)
$$
(The pattern is three plus signs and a minus.  It doesn't matter where
one puts the minus sign.)

\PROOF
It suffices to prove the inequality without the absolute value signs, as
one sees by
reversing the signs of the
$B$-outcomes.
By assumption there exists a measure $\hat\mu$ on the sample
space $\hat\Omega$ of quadruples $(i i' j j')$ whose marginals
agree with $\mu$ on each $(\alfa, \beta)$ pair.
Therefore
$$
\eqalign{
 X(a,b) = &\sum_{i j} i\cdot j\cdot \mu(a_i\cap b_j\,|\, a\cap b)\cr
       = & \sum_{i i' j j'} i\cdot j\cdot
 \hat\mu(i i' j j')
 \ ,
}
$$
with
similar formulas for $X(a',b)$, $X(a,b')$ and $X(a',b')$.
But for any of the possible values of $i,i',j,j'$ we have
$ ij+i'j+ij'-i'j' = (i+i')j + (i-i')j' \le 2$,
since one of the two parentheses must vanish
in every case.
The weighted average with respect to $\hat\mu$
of this combination of $i$'s and $j$'s
is therefore also less
than or equal to 2; hence
$$
   X(a,b) + X(a', b) + X(a, b') - X(a',b')
   =
    \sum_{i i' j j'} (ij+i'j+ij'-i'j') \ \hat\mu(i i' j j')
  \le 2
  \ .
$$
\QED

\subsection{Quantal case}
At level two we have the same setup as before:
a sample space $\Omega$
including setting events $a$, $a'$, $b$, $b'$, {\it etc.};
and we use the same notation, 
in particular $\alfa = a$ or $a'$
and $\beta = b$ or $b'$.
But now we have
on $\Omega$
a {\it quantal} measure $\mu$
and the associated decoherence functional $D$.
We 
are considering 
the situation in which
the events $\alfa_i\cap\beta_j$
correspond to the readings (and settings) of macroscopic instruments,
and so, as announced earlier, we will assume that
the quantal measure $\mu$ of
any one of these macroscopic
``instrument events''
can be interpreted as
an experimental probability ({\it i.e.\ }a frequency).
Having done so,
we can form {\it conditional probabilities} in the standard manner,
as illustrated by the definition of the
$p(\alfa_i,\beta_j)$ in equation (7) below.
The correlators $X(\alfa,\beta)$ are then definable
exactly as in the classical case [equation (4)].
(Notice that we have not attempted to extend the notion of conditional
probability outside the setting of classical (level 1) measure theory.
To our knowledge, there is, unfortunately, no established notion
of ``conditional quantal measure'' or ``conditional decoherence
functional'', of which classical conditional probability would be a
special case.)\footnote{$^\flat$}
{The need for a quantal generalization of conditional probability arises
 in the following only because the experimental probabilities we work
 with are conditioned on specific instrument settings.  For present
 purposes, it thus would not arise at all in the alternative,
 ``minimalist approach'' mentioned earlier.  However, even in such a
 framework, the need would return as soon as one had to condition on
 the specific {\it results} of observations or other processes.}

Notice that the identification of $\mu(Y)=D(Y;Y)$ as
a probability-{\it qua}-frequency
is only consistent
over the whole algebra of instrument events $Y$
if we assume that 
neither
distinct instrument settings
nor 
distinct outcomes for given settings
interfere 
with one another.\footnote{$^\star$}
%
%% RDS: we explained this already ({\it i.e.\ }they ``decohere'').\footnote{$^\dagger$}
%
{This consistency condition is what gives the ``consistent histories''
 interpretation its name.  But there, it is raised to the level of a
 principle.}
In other words, we must assume that
$$
  \QI{D}{\alfa_i \cap \beta_j}{\alfa_k \cap \beta_l}
  = \mu(\alfa_i \cap \beta_j) \delta_{ik} \delta_{jl} \, ,
 \eqno(6)
$$
$\forall \alpha,\beta$;
and 
we also assume 
that all remaining such off-diagonal values of $D$ vanish,
for example,
$\QI{D}{a_i \cap b_j}{a'_{k} \cap b_l}=0$.

\DEFINITION We denote as the {\it experimental probabilities} the
 sixteen numbers,
 $$
   p(\alfa_i,\beta_j)
   =
  \mu(\alfa_i\cap\beta_j | \alfa \cap \beta)
   =
  {\mu(\alfa_i\cap\beta_j) \over \mu(\alfa \cap \beta)}
  \ .
  \eqno(7)
$$

\DEFINITION
The experimental probabilities $p(\alfa_i,\beta_j)$
{\it admit a joint quantal measure}
iff
there exists a decoherence functional $\hat{D}$  on $\hat\Omega$
such that its {\it marginals} agree
with (7)
for each of the four $(\alfa, \beta)$ pairs
({\it i.e.\ }for each of the four possible instrument settings):
$$
 \hat{D}_{ab}(ij;kl)
 \ideq
 \sum_{i'j'k'l'} \hat{D}(i i' j j' \, ;\, k k' l l')
 =
 p(a_i, b_j) \delta_{ik} \delta_{jl}
 \eqno(8\hbox{$ab$})
$$
%
% $$
%  \sum_{i'j'k'l'} \hat{D}(i i' j j' \, ;\, k k' l l') =
%  {D(a_i\cap b_j; a_k \cap b_l) \over \mu(a \cap b)} =
%  p(a_i, b_j) \delta_{ik} \delta_{jl}
%  \eqno(E::quantummarginals)
% $$
%
for $(\alfa, \beta) = (a, b)$;
$$
 \hat{D}_{a'b}(i'j;k'l)
 \ideq
 \sum_{ij'kl'} \hat{D}(i i' j j' \, ;\, k k' l l')
 =
 p(a'_{i'}, b_j) \delta_{i'k'} \delta_{jl}
 \eqno(8\hbox{$a'b$})
$$
%
% $$
%  \sum_{ij'kl'} \hat{D}(i i' j j' \, ;\, k k' l l') =
%  {D(a'_{i'}\cap b_j; a'_{k'} \cap b_l) \over \mu(a' \cap b)} =
%  p(a'_{i'}, b_j) \delta_{i'k'} \delta_{jl}
% $$
%
for $(\alfa,\beta)=(a',b)$;
and similarly for
$(\alfa,\beta)=(a,b')$
and
$(\alfa,\beta)=(a',b')$.
%
% ({\it cf.\ }(E::joint).)

\REMARK
Our use of the word ``marginals'' here is in obvious analogy to its use
in classical measure theory, where, given that
$\Omega=\Omega_1\times\Omega_2$, a marginal probability distribution on
$\Omega_2$ is one induced from $\Omega$ by summing over $\Omega_1$.
Similarly here, the joint decoherence functional $\hat{D}$ on
$\hat\Omega$ induces marginal decoherence functionals
$\hat{D}_{\alfa\beta}$ (and hence marginal quantal measures
$\hat\mu_{\alfa\beta}$) on all the $(\alfa,\beta)$ pairs, as illustrated
in 
(8\hbox{$ab$}) and
(8\hbox{$a'b$}).

Observe that the matching-conditions (8) on the
marginals require more than just agreement with the 16 probabilities
$p(\alfa_i,\beta_j)$.  They also entail the vanishing of the 24
off-diagonal elements $\hat{D}_{\alfa\beta}(ij;kl)$ with $i\not=k$ or
$j\not=l$.
Notice on the other hand, that they do not
refer to any marginals that would involve interference between distinct
instrument settings.

We will see that the Tsirel'son inequality follows from the existence of
such
a joint quantal measure.  However, in order to demonstrate this, we will
need
to apply to $\hat{D}$ a certain basic construction
via which any strongly
positive decoherence functional
gives rise to
a Hilbert space
[14] [15].

\subsection{Hilbert space from (strongly positive) quantal measure}
Consider the vector space $\H_1$
which consists of all
formal linear combinations of the sixteen four-bit strings,
$(i \, i' \, j \, j')$, $i,i',j,j' = \pm1$.
Let $[ii'jj']$ denote a general basis vector of $\H_1$.
That $\hat{D}$ is strongly positive means that
it induces a (possibly degenerate) Hermitian inner product on $\H_1$
given by:
$$
  \bra [ii'jj'] , [kk'll'] \ket  = \hat{D}(ii'jj' ; kk'll')
%%%  ( [ii'jj'] , [kk'll'] )  =    \hat{D}(ii'jj' ; kk'll')
  \ .
$$
In general, $\H_1$ is not a Hilbert space because it contains
vectors with zero norm.
To form a true Hilbert space $\H$
take the quotient of $\H_1$ by the vector subspace $\H_0$ of
zero norm states: $\H=\H_1/\H_0$.
Denote by $|i i' j j'\rangle$
the vector in $\H$ that corresponds to
$[ii'jj']\in\H_1$.
(Regarding members of $\H_1/\H_0$ as equivalence classes, we can
 describe $|i i' j j'\rangle$ as the set of vectors in $\H_1$ that
 differ from $[ii'jj']$ by vectors of zero norm.)
Plainly, the vectors $|i i' j j'\rangle$ span $\H$;
and we have\footnote{$^\flat$}
{For related observations see [20] [21]
[22].}
$$
   \hat{D}(i \, i' \, j \, j' ; k \, k' \, l \, l')
  = \langle i \, i' \, j \, j' \vert k \, k' \, l \, l'\rangle  \ .
  \eqno(9)
$$
This relationship will let us convert the correlators $X(\alfa,\beta)$
into inner products of vectors in $\H$, the key step in our proof of
Theorem \thmB.

\subsection{Correlators as inner products in Hilbert space: proof of Theorem
2}
In this subsection, we state and prove our main result as a theorem.  We
assume that the experimental probabilities admit a joint quantal measure
given by the decoherence functional $\hat{D}$ on $\hat\Omega$,
as specified in equations (8),
and we denote by $\hat\mu$ the corresponding
generalized measure given by the diagonal elements of $\hat{D}$, as in
equation (2).

\LEMMA{\lemmaA}
Let $|a\ket\in\H$ be defined by
$$
  |a\ket = \sum_{i i' j j'} i \cdot |i i' j j'\ket     \eqno(10)
$$
and similarly for $|a'\ket$, $|b\ket$ and $|b'\ket$.
Then
$\braket{a}{a}$ = $\braket{b}{b}$ = $\braket{a'}{a'}$ = $\braket{b'}{b'}$ =
1,
and for any of the four possible pairings of $\alfa=a,a'$ with $\beta=b,b'$,
we have
$$
   \braket{\alfa}{\beta}
   =
   X(\alfa,\beta)
   \ideq
   \sum_{ij} i \cdot j \cdot p(\alfa_i, \beta_j)
   \eqno(11)
$$

\PROOF
We give the proof for $\alfa=a$, $\beta=b$,
the other three cases being strictly analogous.
%
%% We have defined $X(a,b)$ as before in equation (E::def-X).
% $$
%  X(a, b)
%  \equiv
%  \sum_{ij} i \cdot j \cdot p(a_i, b_j)
%  =
%  \sum_{ij} i \cdot j \cdot \mu(a_i \cap b_j | a \cap b)
%  \eqno(E::Xq)
% $$
%
Using equations (8), (9) and (10),
we replace the sum over
diagonal terms in $X(a, b)$ by the full sum:
$$
\eqalign{
  X(a,b)
  =& \sum_{ij}   i \cdot j \cdot p(a_i,b_j)                        \cr
  =& \sum_{ijkl} i \cdot l \cdot p(a_i,b_j) \delta_{ik}\delta_{jl} \cr
  =& \sum_{ijkl} i \cdot l \cdot \hat{D}_{ab}(ij ; kl)        \cr
  =& \sum_{ijkl} i \cdot l \cdot
      \sum_{i'j'k'l'} \hat{D}(ii'jj'\,;\, kk'll')                  \cr
  =& \sum_{ii'jj'kk'll'} i \cdot l \cdot \hat{D}(ii'jj' ; kk'll')  \cr
  =& \sum_{ii'jj'kk'll'} i \cdot l \cdot \braket{ii'jj'}{kk'll'}  \cr
  =& \braket{a}{b}
  \ .}
$$
%
% $$
% \eqalign{
%   X(a,b)
%   =& \sum_{ij} i\cdot j\cdot
%       {D(a_i\cap b_j \,;\, a_i \cap b_j) \over \mu(a \cap b)} \cr
%   =& \sum_{ijkl} i\cdot l\cdot
%       {D(a_i\cap b_j \,;\, a_k \cap b_l) \over \mu(a \cap b)} \cr
%   =&  \sum_{ijkl} i\cdot l\cdot \sum_{i'j'k'l'}
%           \hat{D}(ii'jj'\,;\, kk'll')                   \cr
%   =&  \sum_{ii'jj'kk'll'} i \cdot l\cdot \hat{D}(ii'jj' ; kk'll')\cr
%   =&  \braket{a}{b}
%   \ .}
% $$
%
We must also prove that the vectors $|\alfa\ket$, $|\beta\ket$ have unit
norm. Let us prove for example that $\braket{a}{a}=1$.
To that end,
define the vectors
$$
\eqalign{
  |a\pm\rangle &\equiv  \sum_{i'jj'}\,|\,\pm1 i'jj'\rangle\cr
}
$$
and note that
$\braket{a+}{a-}=0$ by 
(8\hbox{$ab$}) 
with $i=+1$, $k=-1$.
Then
$$
   |a\rangle  = |a+\rangle - |a-\rangle
$$
and we have
$$
\eqalign{
\langle{}a|a\rangle =
& \langle{}a\!+\!|a+\rangle + \langle{}a\!-\!|a-\rangle
 - \langle{}a\!+\!|a-\rangle - \langle{}a\!-\!|a+\rangle\cr
= & \langle{}a\!+\!|a+\rangle + \langle{}a\!-\!|a-\rangle
 + \langle{}a\!+\!|a-\rangle + \langle{}a\!-\!|a+\rangle \cr}
$$
This last line is $\hat{D}(\hat{\Omega};\hat{\Omega})$,
which is 1 by our assumption of normalization.
\QED

We are now ready to prove our main result, that any set of experimental
probabilities which admits a joint quantal measure must respect the
Tsirel'son inequality:

\THEOREM{\thmB}
If there
exists a strongly positive joint decoherence functional $\hat{D}$
on $\hat\Omega$
whose marginals agree with $D$
--- meaning equations (8) hold ---
then
$$
  | \; X(a,b) + X(a',b) +  X(a,b') - X(a',b') \; |  \le  2\sqrt 2  \ .
  \eqno(12)
$$

\REMARK
Instead of saying that the marginals ``agree with $D$'', we could
equally well have said that 
they ``are diagonal and 
yield the experimental probabilities $p(\alfa_i,\beta_j)$''.  
This expresses the theorem in a more self-contained form.

\PROOF
As before, it suffices to prove (12) without the absolute value signs.
Write
$$
   Q \equiv  X(a,b) + X(a',b) +  X(a,b') - X(a',b')
\eqno(13)
$$
which has a ``logical'' maximum value of 4.
%
% (The minus sign can once again, as in the CHSHB inequality, go anywhere in
% $Q$.  The form of the calculations to follow remains essentially the same.)
%
By the previous lemma, we have
$$
\eqalign{
  Q &= \braket{a}{b} + \braket{a'}{b} +  \braket{a}{b'} - \braket{a'}{b'}
 \cr &= ( \langle{}a| + \langle{}a'|) \; |b\rangle
    +  ( \langle{}a| - \langle{}a'|) \; |b'\rangle   \ .
}
 \eqno(14)
$$
Since $|b\rangle$ and $|b'\rangle$ are unit vectors,
$Q$ is maximized when $|b\rangle$ is parallel to $|a\rangle+|a'\rangle$
and $|b'\rangle$ is parallel to $|a\rangle - |a'\rangle$.
Hence
$$
  Q \le
  \Vert |a\rangle + |a'\rangle \Vert
  + \Vert |a\rangle - |a'\rangle \Vert  \ ,
$$
whence $Q\le 2\sqrt{2}$ by the following simple lemma.  \QED

\LEMMA{\lemmaB}
If $u$ and $v$ are vectors of unit length
then $||u+v||+||u-v|| \le \sqrt{8}=2\sqrt{2}$.

\PROOF
Let $S=||u+v||+||u-v||$ and write $\xi=\Re\bra{u}|v\ket$.
Then $||u\pm v||^2 = \braket{u\pm v}{u\pm v}=(1+1\pm2\xi)=2\pm2\xi$.
Hence
$$
\eqalign{
S^2 &= ||u+v||^2 + ||u-v||^2 + 2 ||u-v|| \, ||u+v|| \cr
    &= (2+2\xi) + (2-2\xi) + 2 \sqrt{(2+2\xi)(2-2\xi)} \cr
    &= 4+2 \sqrt{4-4\xi^2}                              \cr
    &\le4+2\sqrt{4} = 8                                 \cr
    &\implies S \le \sqrt{8} \ .                        \cr
}
$$
% Finally, note that $Q$ is also bounded below by $-2\sqrt{2}$, as can be seen
% by redefining the signs of the outcomes.
\QED

\subsection{An example: saturating the bound}
To illustrate some of the above,
consider the familiar quantum mechanical
setup leading to maximal violation of Bell's inequalities (5),
which is known to produce
a system of experimental probabilities that
saturates the Tsirel'son bound (12).
The mathematics involved in this situation
produces a joint decoherence functional that also is on the boundary of
the convex set of strongly positive decoherence functionals.
We can
use this to
conclude
that by itself,
weak positivity of the quantal measure
({\it i.e.\ }the condition $D(X;X)\ge0$ on the decoherence functional)
is insufficient to
imply the bound (12).

We assume we have two spin-half particles in a singlet state and
each particle heads off to either region $A$ or region $B$ where
Alya and Bai, respectively,  are waiting to make measurements
on the particles. Alya sets her apparatus to measure the spin in
directions $\bfa$ or $\bfa'$ and Bai in directions $\bfb$ or $\bfb'$ where
$\bfa$, $\bfa'$, $\bfb$ and $\bfb'$ are now unit vectors in three
dimensional
space satisfying
%
% $$ \eqalign{
% \bfa\cdot\bfa' &= \bfb\cdot \bfb' = 0\cr
% \bfa\cdot\bfb &= \bfa\cdot \bfb' = - \bfa'\cdot \bfb = \bfa'\cdot \bfb' =
% \frac{1}{\sqrt{2}}.
% \cr}
% $$
%
$$ \eqalign{
\bfa\cdot\bfa' &= \bfb\cdot \bfb' = 0\cr
\bfa\cdot\bfb &= \bfa\cdot \bfb' = \bfa'\cdot \bfb = -\bfa'\cdot \bfb' =
\frac{1}{\sqrt{2}} \ .
\cr}
$$

It is interesting that when calculating the quantity $Q$ as
given by ordinary quantum mechanics in this setup we obtain
%
% $$
% \eqalign{
%   & X(\bfa, \bfb) - X(\bfa', \bfb) +  X(\bfa, \bfb') + X(\bfa', \bfb') \cr
%   &=  (\bfa- \bfa') \cdot \bfb  +  (\bfa + \bfa') \cdot \bfb', \cr}
% $$
%
$$
\eqalign{
  & X(\bfa, \bfb) + X(\bfa', \bfb) +  X(\bfa, \bfb') - X(\bfa', \bfb') \cr
  &=  (\bfa + \bfa') \cdot \bfb  +  (\bfa - \bfa') \cdot \bfb', \cr}
$$
which is exactly the same expression (14) as arose in the general
proof of (12), only here we have ordinary vectors in $\Reals^3$
instead of vectors in Hilbert space.

In the EPRB setup we have a 4-dimensional Hilbert space ${\H}$
which is a tensor product of two qubit Hilbert spaces ${\H}_A$
and ${\H}_B$, and $|\psi\rangle = (|\uparrow\rangle_A |\downarrow\rangle_B
- |\downarrow\rangle_A|\uparrow\rangle_B)/\sqrt{2}$ is the singlet state.
On ${\H}_A$ we have Pauli matrices ${\bm\sigma}$
and
on ${\H}_B$ we have Pauli matrices ${\bm\rho}$,
from which we can form projection operators.

We will form the decoherence functional from the expectation
value in the singlet state of strings of
projectors onto the several values of the two spins in the
four directions, ${\bm a, a', b, b'}$.
Specifically, let us
form the decoherence functional using the strings of projectors
appropriate to the results:
``Alya finds the spin to be $i \hbar/2$ in the ${\bm a}$ direction
{\it and then } $i'\hbar/2$ in the ${\bm a'}$ direction,
and
Bai finds $j\hbar/2$ in the ${\bm b}$ direction
{\it and then} $j'\hbar/2$ in the ${\bm b'}$ direction.''
For this we need the projectors
$$
\eqalign{
  P^a_i &\equiv \frac{1}{2}( 1 + i\, {\bm a}\cdot \bm{\sigma}) \cr
  P^{a'}_{i'} &\equiv \frac{1}{2}( 1 + i'\, {\bm a'}\cdot\bm{ \sigma}) \cr
  P^b_j &\equiv \frac{1}{2}( 1 + j \,{\bm b}\cdot \bm{\rho}) \cr
  P^{b'}_{j'} &\equiv \frac{1}{2}( 1 + j'\, {\bm b'}\cdot \bm{\rho}) \,.
\cr}
$$

{}From these
and the initial
%% RDS: single particle
singlet state $|\psi\rangle$
we can construct the following decoherence functional that
is strongly positive and decoheres on all $(\alfa,\beta)$ pairs:
%
% $$
%  \hat D(i i' j j' \,;\, k k' l l')
%  = \langle{}\psi| P^a_i P^{a'}_{i'} P^b_j P^{b'}_{j'}
%                  P^{b'}_{l'} P^b_l P^{a'}_{k'} P^a_{k}
%   |\psi\rangle
% \ .
% \eqno(E::dcf-1)
% $$
%
$$
 \hat D(i i' j j' \,;\, k k' l l')
 = \langle{}\psi| P^a_k P^{a'}_{k'} P^b_l P^{b'}_{l'}
                 P^{b'}_{j'} P^b_j P^{a'}_{i'} P^a_{i}
  |\psi\rangle
\ .
\eqno(15)
$$
(This is just a decoherence functional in the sense of [19],
 evaluated on the coarse-grained histories represented by
 $P^a_i P^{a'}_{i'} P^b_j P^{b'}_{j'}$,  with initial state $|\psi\rangle$.)

The decoherence functional of 
(15)
will do the job, but there's a nicer, more symmetric form that
will also work, where the order of the ${\bm{a}}$ and ${\bm{a'}}$
measurements is symmetrized and similarly for ${\bm{b}}$ and
${\bm{b'}}$:
%
% $$ \eqalign{
% & \hat D_{sym}(ii'jj'\,;\, kk'll')\cr
% =& \frac{1}{16}\langle\psi|
% (P^a_i P^{a'}_{i'} + P^{a'}_{i'}P^a_i) (P^b_j P^{b'}_{j'}
% + P^{b'}_{j'}P^b_j)
%  (P^b_l P^{b'}_{l'} + P^{b'}_{l'}P^b_l)
% (P^a_k P^{a'}_{k'} + P^{a'}_{k'}P^a_k)
% |\psi\rangle . \cr}
% \eqno(E::dcf-2)
% $$
%
$$ \eqalign{
& \hat D_{sym}(ii'jj'\,;\, kk'll')\cr
=& \frac{1}{16}\langle\psi|
(P^a_k P^{a'}_{k'} + P^{a'}_{k'}P^a_k) (P^b_l P^{b'}_{l'}
+ P^{b'}_{l'}P^b_l)
 (P^b_j P^{b'}_{j'} + P^{b'}_{j'}P^b_j)
(P^a_i P^{a'}_{i'} + P^{a'}_{i'}P^a_i)
|\psi\rangle . \cr}
\eqno(16)
$$
Some simple $\sigma$-matrix algebra, using
$({\bm \sigma} + {\bm \rho}) |\psi\rangle =0$ and
$\langle\psi|{\bm \sigma}|\psi\rangle = 0$,
because $|\psi\rangle$ is the singlet, gives
$$ \eqalign{
  &256\, \hat D_{sym}(ii'jj'\,;\, kk'll')\cr
  = &(1 + ik + i'k')(1+ jl + j'l')\cr
  + &(ik' - i'k)(jl' - j'l)\cr
  - & \frac{1}{\sqrt{2}}(i + k)(j + l + j' + l')\cr
  + & \frac{1}{\sqrt{2}}(i' + k')(j + l - j' - l').
  \cr}
  \eqno(17)
$$
One can easily verify that $|Q|=2\sqrt2$ with these numbers.

The $16\times16$ matrix $\hat{D}_{sym}$ has 12 zero eigenvalues and so the
Hilbert space that one constructs from it is four-dimensional,
as one would expect
for a pair of spin-$\half$ particles
[15].
The existence of null directions also means
that $\hat\mu$ is
verging on violating strong
positivity,
the matrix $\hat{D}$ being only positive {\it semi}-definite.

\subsection{Realizing the joint sample space}

The construction we have just employed can be made more vivid by
relating it to a gedankenexperiment in which the 16 ``outcomes''
comprising
the
sample space $\hat\Omega$ correspond to actual trajectories of physical
particles.  The expression (15) can be interpreted in terms of
Stern-Gerlach devices for
silver atoms
(or perhaps more conveniently, in
terms of photon trajectories and interferometers, {\it cf.\ }the
setup in [23].)  An outcome of a spin measurement then
amounts to
a silver atom's
emerging in either the upper or lower beam.
However, suppose that we don't ``look at'' the
silver atom,
but instead
send it through a reversed magnetic field designed to recombine the
two beams, as if they had never been split apart at all.  We can then
pass
it
through a {\it second} Stern-Gerlach analyzer
which again splits the beam into two, {\it etc.} If we
concatenate
two analyzers this
way in region $A$,
and two more in region $B$,
then we naturally partition the
full history space into 16 subsets depending on which beams the
silver atoms
traverse in their respective analyzers.  
In this way 
the elements of $\hat\Omega$ are realized as 
actual sets of histories that
all pertain to a {\it single} experimental setup.
It follows that {\it any
histories formulation}\footnote{$^\star$}
{Any formulation, that is, for which the
silver atoms
 are part of the kinematics (or ``ontology'') and
 trace out continuous
 worldlines in spacetime.  With discontinuous trajectories,
 the silver atoms
 might be present in both beams, and an event like
 ``silver atom
 in upper
 beam in first analyzer'' would not be well defined.  It seems that
 something like this would actually occur in models such as that of
 [24].}
{\it must induce in this manner a quantal measure on}
$\hat\Omega$;
and the
matrix-element
(15) that we wrote
down before is just
the
algebraic expression of this  measure.

Of course, the mere fact that the measure $\hat\mu$ is well defined
does not yet tell us what will happen if we do ``look at'' the
particles.  To make contact with the experimental probabilities
$p(\alfa_i,\beta_j)$, we must assume further that if we do choose to
look, then the measure induced thereby on ``us'' directly reflects
the measure $\hat\mu$ on the space $\hat\Omega$ of ``microscopic''
alternatives.
That is, we must assume that ``looking'' at a beam {\it merely reveals}
the corresponding value of $\hat\mu$.
(Note in this connection that 
 events in
 distinct spatial locations at a given time
 always decohere in unitary quantum mechanics.)

\section{IV. Relation to the ``screening off'' causality condition}%
\subsection{Classical case}
The condition of {\it screening off} on the
classical measure $\mu$
asserts that
events in causally unrelated regions $A$, $B$ of spacetime become
independent
(``screened off'' from one another)
when one conditions on a complete specification $c$ of the
history in
the region $C=J^-(A)\cap J^-(B)$,
the mutual causal past\footnote{$^\dagger$}
{This is a strong form of the screening off condition, as it excludes in
 particular ``primordial correlations''.  A less restrictive condition,
 depending on the context, would locate $c$ in the {\it union} of the
 (exclusive) pasts of $A$ and $B$.  For details see [9].}
of $A$ and $B$.
The logic
underlying
this condition is that any correlation between
spacelike separated variables
must arise
entirely
from their separate correlations with some ``common cause'' in
their mutual past, and therefore
must disappear once
full information
about the past is
given.\footnote{$^\flat$}
{This is not the only way to construe the ``principle of common cause'',
 but it is the one adopted in all discussions of the Bell inequalities
 known to us.}

Specialized to our situation, this screening off condition yields
for all $\alfa\in M_A$, $\beta\in M_B$ and $i,j = \pm 1$,
$$
 \mu(\alfa_i \cap {\beta_j} | c)
 = \mu(\alfa_i | c)  \,
   \mu(\beta_j | c)
  \ ,
  \eqno{(18)}
$$
where
$c$ is any subset of $\Omega$ defined by a completely
fine grained specification of
the history in $C=J^-(A)\cap J^-(B)$.
Similarly (or just by summing (18) on $i,j$), we have
$$
  \mu(\alfa \cap \beta |c) = \mu(\alfa|c) \mu(\beta|c),
  \eqno{(19)}
$$
so that the ``setting event'' $\alfa$ is screened off from the setting event
$\beta$.
Dividing (18) by (19) yields an equation which,
we claim,
can be written
as
$$
  \mu(\alfa_i \cap \beta_j | \alfa \cap \beta \cap c)=
  \mu(\alfa_i| \alfa \cap c) \mu(\beta_j | \beta \cap c) \ .
 \eqno{(20)}
$$
\extraspace
This follows from noting that
$$ \eqalign{
\mu(\alfa_i \cap \beta_j | \alfa \cap \beta \cap c)
&=
{\mu(\alfa_i \cap \beta_j \cap \alfa \cap \beta \cap c)\over \mu(\alfa \cap
 \beta \cap c)}  \cr
&=
{\mu(\alfa_i \cap \beta_j \cap c) \over \mu(\alfa \cap \beta \cap c)}
        \qquad\quad [{\rm since}\ \alfa_i\subset \alfa,\ \beta_j\subset
\beta]    \cr
& =
{\mu(\alfa_i \cap \beta_j \cap c)/\mu(c) \over \mu(\alfa \cap \beta \cap c)
/ \mu(c)}  \cr
&=
{\mu(\alfa_i \cap \beta_j | c) \over \mu(\alfa \cap \beta | c)}
}
$$
and
$$ \eqalign{
\mu(\alfa_i|\alfa \cap c)
&=
{\mu(\alfa_i \cap \alfa \cap c) \over \mu(\alfa \cap c)}   \cr
&=
{\mu(\alfa_i        \cap c) \over \mu(\alfa \cap c)}   \cr
&=
{\mu(\alfa_i \cap c) / \mu(c) \over \mu(\alfa \cap c) / \mu(c)}    \cr
&=
{\mu(\alfa_i|c) \over \mu(\alfa|c)},
}
$$
and similarly for $\mu(\beta_j|\beta \cap c)$.
[In these calculations, one is in effect ``conditioning in stages''
 and recognizing that
 $\mu((x|y)|z)=\mu(x|y \cap z)$, where
 $\mu((x|y)|z):=\mu(x \cap y|z)/\mu(y|z)$.]
Observe that, in order for the conditional probabilities appearing in
(18)--(20) to be defined, none of the measures
$\mu(c)$, $\mu(\alfa \cap c)$, $\mu(\beta \cap c)$, $\mu(\alfa \cap \beta
\cap c)$
can vanish.  Accordingly, (20) is only valid with this reservation.

\sesquispace

At this point,
we need
to formalize the idea that the instrument
settings are ``chosen freely''.  To that end, we will assume that,
with respect to the measure $\mu$,
 and
for all $\alfa\in M_A$ and $\beta\in M_B$,
the ``setting events'' $\alfa$ and $\beta$ 
are independent of any\footnote{$^\star$}
{This is a rather drastic form of setting-independence.
 It would have been possible to include other events in the past on
 which the settings depended without affecting the main points of the
 argument, as in [10].}
events in $C$.
In the presence of screening off, this implies that the event
``$\alfa$ and $\beta$'' is also independent of any event in $C$:
$$
   \mu(\alfa \cap \beta \cap c) = \mu(\alfa \cap \beta) \mu(c) \ .
 \eqno{(21)}
$$
(It also implies that $\mu(\alfa\cap \beta)=\mu(\alfa)\mu(\beta)$,
so that, in the presence
of screening off, the setting events are strictly independent of one
another.)
The
formal
derivation of (21) goes as follows.
Our assumption of ``setting-independence'' says
that
$$
   \mu(\alfa|c)=\mu(\alfa) \ \ (\hbox{and similarly for } \beta) \ .
   \eqno(22)
$$
Putting this together with (19), we
obtain
$\mu(\alfa \cap \beta|c)$
=
$\mu(\alfa|c)\mu(\beta|c)$
=
$\mu(\alfa)\mu(\beta)$,
whence
$\mu(\alfa \cap \beta \cap c)=\mu(\alfa)\mu(\beta)\mu(c)$,
whence
$\mu(\alfa \cap \beta)=\mu(\alfa)\mu(\beta)$ by summing on $c$.
Comparing the first and last equations yields
$\mu(\alfa\cap \beta | c)=\mu(\alfa\cap \beta)$, which is
(21).
Note finally that we can assume without loss of generality that
$\mu(c)>0$ for all fine-grained specifications $c$ of $C$ (otherwise
simply
omit $c$ from $\Omega$).
Then, having just demonstrated that
$\mu(\alfa\cap \beta\cap c)=\mu(\alfa)\mu(\beta)\mu(c)$,
we conclude that
$\mu(\alfa\cap \beta\cap c)$
never vanishes (unless we can't do the experiment at all!); hence
equation (20) becomes valid unreservedly.

It is well known that the screening off condition leads to the
CHSHB inequality [10].
This follows from a
result of Fine [4] according to which
the existence of a joint distribution $\hat\mu$ on $\hat\Omega$
is equivalent\footnote{$^\dagger$}
{Fine's treatment appears to rely tacitly on the ``non-contextuality''
 assumption that
 settings of the remote instrument cannot affect local results.
 The condition that he invokes is not actually screening off as such,
 but what he calls ``factorizability'', a condition which, as he words
 it, seems to be
 ambiguous between two formulations, the first of which (corresponding
 to our equation (20)) could be written in our notation as
 $\mu(\alfa_i\cap \beta_j | \alfbetc) = \mu(\alfa_i|\alfa\cap c)
\mu(\beta_j|\beta\cap c)$,
 and the second of which would be
 $\mu(\alfa_i\cap \beta_j | \alfbetc) = \mu(\alfa_i| \alfbetc) \mu(\beta_j |
\alfbetc)$.
 In these expressions, however, ``conditioning'' on $\alfa$ 
(for example)
 merely
 means that the instrument at $A$ is set to $\alfa$.  
 Fine avoids attributing probabilities to instrument settings, 
 in contrast to the approach we have adopted in this paper; 
 his is the ``minimalist approach'' 
 mooted at the beginning of Section III.} 
%% {\it cf.\ }the discussion at the end of this subsection.}
%
to
screening off.
More formally, let us say that a system of experimental probabilities
$p(\alfa_i,\beta_j)$
{\it admits a
classical
screening off model}
if one can find a sample space $\Omega$ and a measure $\mu$ thereon
obeying (18)
and (21),
and such that
$\mu(\alfa_{i}\cap\beta_{j} | \alfa\cap\beta) = p(\alfa_i,\beta_j)$.
Then

\LEMMA{\csojm}
A system of experimental probabilities admits a
classical
screening
off model
if and only if it admits a joint probability
distribution $\hat\mu$.

\PROOF
(1) Let  $(\Omega,\mu)$ be a screening off model for some system of
experimental probabilities.
We must demonstrate that
there exists a classical measure, $\hat\mu$ on $\hat\Omega$, such that
$$
\mu(a_i \cap b_j | a \cap b)
  = \sum_{i'j'} \hat\mu(i \, i' \, j \, j')
$$
and similarly for each $(\alfa, \beta)$ pair.
% $\alfa = a$ or $a'$,
% $\beta = b$ or $b'$.
%
In the following,
recall that $c$ ranges over all subsets of $\Omega$ specified by a
complete fine-grained description of $J^-(A) \cap J^-(B)$
for which $\mu(c)>0$.
Recall also that we have assumed that $\Omega$ has finite cardinality.

Since,
by our assumptions, 
$\mu(a\cap c)$ never vanishes,
%% $\mu(a\cap c) = \mu(a)\mu(c)>0$,
$\mu(a_i \,|\, a\cap c)$ is defined,
and
we have
$$
   \sum_{i} \mu(a_i \,|\, a\cap c) = 1  \ ,     \eqno{(23)}
$$
and similarly for $a',b,b'$.
Now make the ``maximal independence ansatz'',
$$
 \hat\mu(i \, i' \, j \, j')
  = \sum_c \mu(a_i\vert a\cap c)        \,
           \mu(a'_{i'}\,|\, a'\cap c) \,
           \mu(b_j\,|\, b\cap c)        \,
           \mu(b'_{j'}\,|\, b'\cap c) \,
           \mu(c)
           \ .                           \eqno{(24)}
$$
We claim that the marginals of $\hat\mu$
agree with
$\mu$ for each
$(\alfa,\beta)$ pair.
For example, take $\alfa=a$, $\beta=b$; then
$$
\eqalign{
  {}& \sum_{i' j'} \hat\mu(ii'jj')                             \cr
    & = \sum_c \mu(a_i \vert a\cap c) \mu(b_j \,|\, b\cap c) \mu(c)
        \qquad [{\rm by}\ (23)]                               \cr
    &= \sum_c \mu(a_i \cap b_j | a \cap b \cap c) \mu(c)
        \qquad\quad [{\rm  by}\ (20)]                          \cr
    &= \sum_c {\mu(a_i \cap b_j \cap c)\over \mu(\abc)} \mu(c) \cr
    &= \sum_c {\mu(a_i \cap b_j \cap c)\over \mu(a \cap b)\mu(c)} \mu(c)
        \qquad [{\rm by}\ (21)]                         \cr
    &= {\sum\limits_c \mu(a_i \cap b_j \cap c) \over \mu(a \cap b)}   \cr
    &= {\mu(a_i \cap b_j) \over \mu(a \cap b)}  \cr
    &= \mu(a_i \cap b_j | a \cap b ) \ .                       \cr }
$$

\noindent
(2) Conversely, suppose there exists a classical measure $\hat\mu$ on
$\hat\Omega$ as described above with marginals that agree with $\mu$ on
each $(\alfa, \beta)$ pair.
% Even if screening off does not hold for
% $\mu$,
A consistent screening
off model can
% still
be found by supposing that there were other events
in the past which were not taken into account in
the original sample space $\Omega$.

We will assume for the purposes of this proof that the original sample
space $\Omega$ contains nothing but the experimental events of interest,
{\it i.e.\ }the settings and outcomes, and in particular contains no
events in the 
mutual
past of $A$ and $B$.  It would have been possible to include past events
of which all the experimental probabilities in $\Omega$ were independent;
no essentially new idea is needed to extend the proofs to this case, but
they
are excluded here for the sake of clarity.

Let $\widetilde\Omega$ be a new sample space whose fine-grained histories
are
those of $\Omega$ with an additional quadruple of binary variables
which we will regard as residing in the mutual past of $A$ and $B$.
(These variables play the role of the past ``causes''
of the experimental outcomes.)
We claim that
there exists a classical measure $\widetilde\mu$ on $\widetilde\Omega$ 
such that the
measure $\widetilde\mu$ agrees with $\mu$ on all of the
experimental events --- and
that
$\widetilde\mu$ satisfies screening off.
To demonstrate this, let us
set, formally,
$$
\widetilde\Omega =
 \SetOf { (h, i , i' , j , j') }
  {h \in \Omega {\rm \ and\ } i, i', j, j' \in \braces{+1, -1}}
$$
and declare by fiat that the quadruple of binary variables
$(i,i',j,j')$
lives in the mutual past of $A$ and $B$.
Then any subset of $\Omega$ can be considered a subset of $\widetilde\Omega$
in the obvious way.
We write $\braces{i i' j j'}$ or $\braces{k k' l l'}$ for the
set of all histories in $\widetilde\Omega$ with those particular values of
the quadruple in the past.

The statement of screening off for this new model is
$$ \eqalign{
   \widetilde\mu(\alfa_i\cap \beta_j \,|\, \braces{ kk'll' } )
 = \widetilde\mu(\alfa_i \,|\, \braces{kk'll'})
   \widetilde\mu(\beta_j \,|\, \braces{kk'll'})
}.
    \eqno{(25)}
$$
We must find a $\widetilde\mu$
which extends $\mu$ and
for which (25) holds.
Note that the experimental
settings are still
required
to be independent of all past events,
which now means $\braces{kk'll'}$.
We make the ansatz
$$
\eqalign{
 \widetilde\mu(a_i \cap b_j \cap \braces{k k' l l'})
 & = \mu(a \cap b)  \,
     \hmu(k k' l l') \, \delta_{i k} \delta_{j l} \cr
 \widetilde\mu(a'_{i'} \cap b_{j} \cap \braces{k k' l l'})
 & = \mu(a' \cap b) \,
     \hmu(k k' l l') \, \delta_{i' k'} \delta_{j l}
}
 \eqno{(26)}
$$
and similarly for the other two $(\alfa,\beta)$ pairs,
$(a,b')$ and $(a',b')$.

Summing one example of (26) over $k,k',l,l'$
yields, with the help of (3),
$$
 \widetilde\mu(a_i\cap b_j)
  =\mu(a \cap b ) \sum_{i' j'} \hmu(ii'jj')
  =\mu(a_i\cap b_j)
  \ .
$$
This shows that the probabilities of the experimental outcomes and
settings are the same for $\mu$ and $\widetilde\mu$.

As required, the settings are also independent of the added past
variables with respect to $\widetilde\mu$, for example:
$$
\eqalign{
 \widetilde\mu(a \cap b \cap \braces{k k' l l'})
 & = \sum_{ij} \widetilde\mu(a_i \cap b_j \cap \braces{k k' l l'}) \cr
 & = \mu(a \cap b)  \, \hmu(k k' l l') \cr
 & = \widetilde\mu(a \cap b)  \, \widetilde\mu(\braces{k k' l l'}).
}
$$

The ansatz (26) also gives,
after simple manipulations,
$$ \eqalign{
 {}&\widetilde\mu(a_i\cap b_j \,|\,\braces{kk'll'})  = \delta_{ik}
\delta_{jl} \widetilde\mu(a \cap b) \cr
 {}&\widetilde\mu(a_i \,|\,\braces{kk'll'})  = \delta_{ik}  \widetilde\mu(a)
\cr
 {}&\widetilde\mu(b_j \,|\,\braces{kk'll'})  = \delta_{jl}  \widetilde\mu(b)
 \ ,}
$$
which implies (25), so the new measure $\widetilde\mu$ 
satisfies screening off.
\QED

The significance of the second part of the lemma is that, given the
existence of the joint probability measure $\hmu$ on $\hOm$, the
observed experimental probabilities can always be explained classically
and causally, in a suitably chosen model. (In the proof of this part of
the lemma, the underlying idea is almost trivial, despite the somewhat
complicated notation that expresses it in this case: if the past
{\it determines} the future, then {\it any} two future events become
independent when the past is conditioned upon.  Screening off is thus
automatic in any ``deterministic'' situation.  The same basic fact
persists for quantal measures and will underlie our proof of Lemma
\qjmso in the next subsection (where the notational complications are
even greater).)

\COROLLARY Screening off $\implies$ the CHSHB inequality.

\REMARK
By conditioning on a given instrumental setup $(\alfa,\beta)$,
we obtain from the overall measure $\mu$ a probability  measure
on the space
$W_{\alfa\beta}=\Omega_\alfa\times\Omega_\beta\times\Omega_C$,
where $\Omega_C$ is the space of all configurations or ``partial
histories'' in the past region $C$.
In this way, we obtain four distinct ``measure spaces'', where by
this phrase we simply mean a sample-space endowed with a measure.
The proof of part (1) of the lemma in effect ``patches'' these
four measure spaces together into a single measure space
$W=\Omega_a\times\Omega_{a'}\times\Omega_b\times\Omega_{b'}\times\Omega_C$,
in a manner reminiscent of a fibre product, with $\Omega_C$ playing the
role of common base space.
The ansatz (24) then produces $\hat\Omega$ (with $\hat\mu$) as the
marginal measure space resulting from $W$ by neglecting $C$.

\subsection{An alternative to screening off?}

This might be
an appropriate
place to comment on the possibility of a different
derivation of the CHSHB inequality, in which an
enhanced locality condition does some of the work done by screening off
in the proof of Lemma \csojm.
Can one, in fact,
demonstrate the existence of a joint decoherence functional without
invoking screening off as such?  This is an interesting question because
it is perhaps not settled that screening off is the true
expression of relativistic causality, even in the classical case
[9] [25].
Here, then, is such an alternative derivation (albeit not as precisely
formulated).

We start from the assumption that instruments at $A$ (resp.\ $B$) respond
only
to certain local variables (``beables'') $\xi_A$ (resp.\ $\xi_B$) defined in
$A$ (resp.\ $B$).  The perfect correlations that arise in the singlet state
then imply that these local variables {\it determine} the response
unambiguously, without any stochastic component (this being the EPR
observation); and we may assume that this is always so, even in examples
such
as that of [6], where the correlations are not perfect.  On this
basis, we immediately acquire our sample space $\hat\Omega$,
% --
parameterized by the values of the local variables $\xi_{A,B}$.

We also need a matching probability measure $\hat\mu$ on $\hat\Omega$.
For this, we must assume that the choice of instrument setting at $A$
--- including the choice of no measurement at all ---
cannot influence local variables at $B$, and vice versa for
settings at $B$.
It follows that we get a well defined (setting independent) probability
distribution on the
variables $\xi_A$, $\xi_B$, and this induces a probability measure on
$\hat\Omega$, whose marginals are obviously the experimental
probabilities, $p(a_i,b_j)$, {\it etc.}

We see that screening off as such was not used.  In its place was the
assumption that instrument settings do not influence the ``hidden
variables''
$\xi_A$ and $\xi_B$.  We tacitly assumed as well, of course, that the
variables $\xi$ do not influence the instrument settings, {\it i.e.\ }that
the
latter were ``free'' in relation to this particular set of microscopic
variables.  Notice also that the derivation in this form did not require us
to
attribute probabilities to instrument settings, except insofar as this would
be one way to make precise their ``freedom'' in respect of the variables
$\xi$.

In the derivation just described, the ``local beables'' $\xi$ provide
a ``material basis'' for the sample space $\hat\Omega$, in the sense that
elements of $\hat\Omega$ represent equivalence classes of histories
determined by the values of those beables.  In contrast,
the proof we gave earlier merely {\it concocted} a space $\hat\Omega$
(and a measure on it),
without
attempting
to identify it with any actual set of
physical
histories
({\it cf.\ }the remarks under ``{Realizing the joint sample space}'',
above.)

\subsection{Quantal case; Proposal for quantal screening off}
We have seen that classically, there is an equivalence between screening
off and the existence of a joint probability measure $\hat\mu$ on all of
the outcomes under consideration, {\it i.e.\ }on $\hat\Omega$.
We would like to prove something similar in the quantum case: that a
suitably generalized screening off condition is equivalent to
the existence of a joint {\it quantal} measure on $\hat\Omega$, all
subject to appropriate conditions of setting independence and
decoherence.  Given our standing assumption of strong positivity, the
Tsirel'son inequality would then follow from quantal screening off.

We will propose a candidate
for a condition of quantum screening off such that,
if there exists a joint decoherence functional
$\hat{D}$
with
the correct marginals,
then a past can be cooked up, just as in the classical case, so that the
resulting
quantum measure $\widetilde\mu$  satisfies the proposed condition.
The converse of this, that the
candidate quantum screening off condition implies
the existence of a strongly positive joint decoherence functional
on $\hat\Omega$
with the correct marginals
remains conjectural for now.  We will, however, prove that
such a decoherence functional on $\hat\Omega$
exists in
the case of ordinary unitary quantum mechanics, and
we will
highlight
the causality assumption that allows the construction.

To motivate the proposed quantal screening off condition, notice that
the {\it classical} screening off condition (18) is equivalent
to
$$
  \mu(\alfa_i\cap \beta_j \cap c) \mu(c)
= \mu(\alfa_i\cap c)\mu({\beta_j}\cap c) \ .
$$
The analogous condition on $D$ is then our proposal for quantum
screening off:%
$$
 D(\alfa_i\cap{\beta_j}\cap c
 \,;\, \bar{\alfa}_k\cap\bar{\beta}_l\cap
 \bar{c}) D(c\,;\, \bar{c})
 =
 D(\alfa_i \cap c
 \,;\,\bar{\alfa}_k \cap \bar{c})
 D({\beta_j}\cap c
 \,;\, \bar{\beta}_l \cap \bar{c})
 \ ,
 \eqno{(27)}  % 27
$$
for all settings $\alfa$, $\bar\alpha$ and $\beta$, $\bar\beta$,
and for all
fully specified pasts
$c$, $\bar{c}$
[as in (18)].
More details and a proof that quantum field theory satisfies this
condition formally will appear in a separate work [25].
(Conditions (27) include matrix elements that are off-diagonal
 in the instrument settings, $\alfa$ and $\beta$.  With the ``minimalist
 approach'', only the equalities with $\alfa=\bar\alfa$ and
 $\beta=\bar\beta$ would be meaningful.)
Notice that if $D$ is completely diagonal, then
(27) reduces to classical screening off.
A variation on (27) asserts (in a shorthand notation)
that\footnote{$^\flat$}
{The pattern might clarify as: 
 $\QI{D}{ijk}{lmn} \QI{D}{i'j'k}{l'm'n} = \QI{D}{i'jk}{l'mn} \QI{D}{ij'k}{lm'n}$}
$$
  \QI{D}{ijc}{\bar{k}\,\bar{l}\,\bar{c}} \;
  \QI{D}{pqc}{\bar{r}\,\bar{s}\,\bar{c}} =
  \QI{D}{pjc}{\bar{r}\,\bar{l}\,\bar{c}} \;
  \QI{D}{iqc}{\bar{k}\,\bar{s}\,\bar{c}}
  \eqno(28)
$$
%% [[RDS  I commented out two lines here that were causing an error:]] 
% $$
%   \QI{D}{ijc}{\bar{l}\,\bar{m}\,\bar{c}}
% \QI{D}{rsc}{\bar{r}\,\bar{s}\,\bar{c}} =
% \QI{D}{rjc}{\bar{r}\,\bar{m}\,\bar{c}}
% \QI{D}{isc}{\bar{l}\,\bar{s}\,\bar{c}}
% $$
where $c$ and $\bar{c}$ are as before, every other index
stands for an event in region $A$ or $B$, and indices appear
in the order: $A$-event, $B$-event, mutual past.
From (28) 
one can deduce
that $D$ decomposes as a product of the form
$$
  \QI{D}{ijc}{\bar{k}\,\bar{l}\,\bar{c}} = \QI{F}{ic}{\bar{k}\,\bar{c}} \;
  \QI{G}{jc}{\bar{l}\,\bar{c}} \ .
  \eqno(29)
$$
% $$
%   \QI{D}{ijc}{\bar{l}\,\bar{m}\,\bar{c}} = \QI{F}{ic}{\bar{l}\,\bar{c}} \;
%   \QI{G}{jc}{\bar{m}\,\bar{c}} \ .
% $$
This formulation carries more information than (27) when
$\QI{D}{c}{\bar{c}}=0$, 
which can happen non-trivially in the quantal case.

% As in the classical case, we will say that
% a system of experimental probabilities
% {\it admits a quantal screening off model}
% if one can find a sample space $\Omega$ and a strongly positive
% decoherence functional $D$ thereon that
% obeys (E::qso),
% and has the correct marginals (E::).
% decoheres on each $(\alfa,\beta)$ pair,
% % and has the correct marginals,
% and correctly reproduces the experimental probabilities
% $p(\alfa_i,\beta_j)=\mu(\alfa_{i}\cap\beta_{j} | \alfa\cap\beta)$.
% %% $p_{\alfa\beta}(i,j)=\mu(\alfa_{i} \cap \beta_{j} | \alfa\cap\beta)$.
% %% $\mu(a'_{i'} \cap b_{j}|a'\cap b)$, etc.

We 
also 
take the quantum condition of setting independence
to be
$$ \eqalign{
  D(\alfa\cap \beta \cap c \,;\, \alfa\cap \beta \cap \bar{c})
  &= D(\alfa\,;\,\alfa)D(\beta\,;\,\beta)D(c\,;\, \bar{c})\cr
  &= \mu(\alfa) \mu(\beta) D(c\,;\, \bar{c}),
}
$$
where
$\alfa\in M_A$, $\beta\in M_B$;
%% $\mu$ is the quantal measure $\mu(x)=D(x;x)$,
$c$ and $\bar{c}$
can be any two events in $C$;
and we
have assumed
that $D$ is diagonal in $\alfa$ and $\beta$.
Finally, recall that all decoherence functionals are assumed by default
to be strongly positive.

In the next lemma, the augmented history space $\widetilde\Omega$ is the
same space as appeared in the proof of Lemma \csojm, part 2.
Also, of course, $\mu$ is the quantal measure on $\Omega$ and $D$ its
associated decoherence functional.

%% RDS: In next lemma do we realy need the decoherence assumption?
%% Probably all we really need is that different *settings* decohere
%% (or no assumption at all in ``minimalist approach'')
%% and then we can reproduce whatever marginals there are, whether or
%% not they're diagonal.

%% RDS: In next lemma what if we assumed only weak positivity?  Would we
%% get a weakly positive mu-hat (ie would we get a q-measure at all)?

\LEMMA{\qjmso}
Let $D$ be a decoherence functional on $\Omega$
that decoheres on $(\alfa,\beta)$ pairs and
such that $\alfa$ and $\beta$ are independent of everything else.
Assume that the induced experimental probabilities
$p(\alfa_i,\beta_j)$
admit a joint quantal measure
in the sense of the definition given above
in equations (7) and (8).
%
% namely a (strongly positive)
% decoherence functional $\hat{D}$ on $\hat\Omega$
% whose marginals $\hat{D}_{\alfa\beta}$ agree
% with $D$ on all $(\alfa,\beta)$ pairs:
% $$
%  \QI{\hat{D}_{\alfa\beta}}{ij}{kl} =
%  \sum_{i'j'k'l'} \hat{D}(i i' j j' \, ;\, k k' l l') =
%  \QI{D}{\alfa_i\cap\beta_j}{\alfa_k\cap\beta_l} / \mu(\alfa\cap\beta)
%  \ .
% $$
% and similarly for the other $(\alfa, \beta)$ pairs.
%
Then there exists
a (strongly positive)
decoherence functional $\widetilde{D}$ on $\widetilde\Omega$
that agrees with $D$ on all pairs of instrument events, and that
satisfies quantum screening off.

\PROOF
As in the classical case,
we assume that $\Omega$ contains no ``irrelevant" events.
We again concoct
extra events $\braces{ii'jj'}$ in the region $C$
that were not taken into
account in $\Omega$.
Our screening off condition for the new
model based on $\widetilde\Omega$ is
$$
\eqalign{
 &\widetilde{D}(\alfa_i\cap{\beta_j}\cap \braces{kk'll'}
 \,;\, \bar{\alfa}_m\cap\bar{\beta}_n\cap
 \braces{pp'qq'}) \; \widetilde{D}(\braces{kk'll'} \,;\, \braces{pp'qq'}) \cr
 &=
 \widetilde{D}(\alfa_i \cap \braces{kk'll'}
 \,;\,\bar{\alfa}_m \cap \braces{pp'qq'})
 \; \widetilde{D}({\beta_j}\cap \braces{kk'll'}
 \,;\, \bar{\beta}_n \cap \braces{pp'qq'})  \ ,
 }
\eqno{(30)}
$$
$\forall$ $\alfa$ and $\beta$.
Define the decoherence functional $\widetilde{D}$ on $\widetilde\Omega$ 
by the equations,
$$
 \eqalign{
  & \QI{\widetilde{D}}{a_i\cap b_j \cap \braces{kk'll'}}{a_m\cap b_n
\cap\braces{pp'qq'}} \cr
  &= \delta_{ik}\delta_{jl} \delta_{mp}\delta_{nq}
  \;
  \mu(a \cap b)
  \;
  \QI{\hat D} {kk'll'}{pp'qq'}
  \ ,}
  \eqno{(31)}
$$
and similarly for the 3 other $(\alfa,\beta)$ pairs,
taking $\widetilde{D}$ 
to vanish when the 
instrument settings are
off-diagonal 
{\it e.g.}
$$
  \QI{\widetilde{D}} {a'_{i'}\cap b_j \cap\braces{kk'll'}}
          {a_m    \cap b_n \cap\braces{pp'qq'}} = 0  \ .
$$
From this, it can be seen that
$\widetilde{D}(\braces{kk'll'} \,;\,
\braces{pp'qq'})=\hat D(kk'll'\,;\,pp'qq')$.

Now, summing, for example, (31) over $k,k',l,l'$ and $p,p',q,q'$
produces
$$
\eqalign{
 {}&\widetilde{D}(a_i\cap b_j  \,;\,  a_m \cap b_n )
  =\mu(a \cap b ) \sum_{k'l'p'q'}
   \hat D(ik'jl'  \,;\, mp'nq') \cr
 &=\mu(a \cap b) p(a_i,b_j) \delta_{im}\delta_{jn} \cr
 &=\mu(a_i\cap b_j) \delta_{im}\delta_{jn} \cr
 &= D(a_i\cap b_j  \,;\,  a_m \cap b_n)
 }
$$
using (8), (7) and (6).
This shows that $\widetilde{D}$ 
takes the same values as $D$ for all pairs
of experimental settings and outcomes.

As required, the setting-events are also independent of the added past
variables with respect to $\widetilde{D}$, for example:
$$
\eqalign{
 \QI{\widetilde{D}}{a \cap b \cap \braces{k k' l l'}}
        {a \cap b \cap \braces{p p' q q'}}
 & = \sum_{ijmn}
   \QI{\widetilde{D}}
      {a_i\cap b_j \cap \braces{kk'll'}}
      {a_m\cap b_n \cap \braces{pp'qq'}} \cr
 & = \mu(a \cap b) \hat D(kk'll' \, ; \, pp'qq') \cr
 & = \widetilde\mu(a \cap b)     \widetilde{D} (\braces{k k' l l'};\braces{p
     p' q q'}) \ .\cr }
$$

The definition of $\widetilde{D}$ also gives
$$ \eqalign{
 {}&\widetilde{D}(a_i \cap \braces{kk'll'} \,;\, a_m \cap \braces{pp'qq'})
=
   \delta_{ik} \delta_{mp}  \widetilde\mu(a) \hat D(kk'll' \,;\, pp'qq') \cr
 {}&\widetilde{D}(b_j \cap \braces{kk'll'} \,;\, b_n \cap \braces{pp'qq'})
=
   \delta_{jl} \delta_{nq} \widetilde\mu(b) \hat D(kk'll'  \,;\, pp'qq')
   \ ,}
$$
which implies (30) 
for $\alfa = \bar{\alfa} = a$
and $\beta = \bar{\beta} = b$.
Similar calculations can be done for every $(\alfa, \beta)$ pair,
and when the instrument settings are
off-diagonal (30) holds trivially as both sides are
zero. 
The new measure $\widetilde\mu$ thus satisfies quantum
screening off.

Finally,
$\widetilde{D}$ is strongly positive because it is essentially just
$\hat D$  which
is strongly positive by assumption.

\QED

We lack a proof of the converse of Lemma \qjmso
(an analogue of part 1 of Lemma \csojm).
But we
{\it can} show
that a strongly
positive decoherence functional on
$\hat\Omega$ with the correct marginals, and which decoheres on
all $(\alfa, \beta)$ pairs, exists in the case of unitary
quantum mechanics
({\it cf.\ }our earlier discussion of concatenated Stern-Gerlach beam
splitters with ``recombiners'', which suggests more generally that
$\hat\Omega$ should be realizable in any ``histories formulation'').

In standard quantum mechanics,
for any measurement $\alfa$ in $A$,
there exist projection operators $P^\alfa_{i}$, $i = \pm1$,
which project onto the subspaces of Hilbert space
associated with the outcomes $\pm 1$ of the measurement.
Plainly,  $P^\alfa_{+1} + P^\alfa_{-1}=1$.
Similarly there
exist operators $P^\beta_{j}$, $j = \pm 1$,
projecting onto the subspaces of
Hilbert space associated with the outcomes $\pm 1$ of
the measurement $\beta$ in $B$.
The standard causality assumption is then
$[P^\alfa_i, P^\beta_j] = 0$.

Given this, it is easy to construct,
analogously to
(15), a joint
decoherence functional on $\hat\Omega$ with the desired properties.
Let
$$
  \hat{D}(ii'jj'; kk'll') = Tr( P^{b'}_{j'} P^b_j P^{a'}_{i'} P^a_i
  \rho_0 P^a_k P^{a'}_{k'} P^b_{l} P^{b'}_{l'})\,,
 \eqno(32)
$$
% $$
%   \hat{D}(ii'jj'; kk'll') = Tr(P^a_i P^{a'}_{i'} P^b_j P^{b'}_{j'}
%   \rho_0 P^{b'}_{l'} P^b_{l} P^{a'}_{k'} P^a_k)\,,
% $$
where $\rho_0$ is the density matrix giving the pre-measurement state of
the particles
and the trace is over particle states.
% (but not those of the analyzers).

\LEMMA{\lsp}
$\hat{D}$ is strongly positive
and has the correct marginals (8).

\PROOF $\hat{D}$ has the canonical form of a decoherence functional of
ordinary unitary quantum mechanics, which is known to be strongly positive
[21].  To show that it has the correct marginals
(8)
on each $(\alfa,\beta)$ pair, let us work out for example the case
$(\alfa,\beta) = (a,b)$:
$$
\eqalign{
\sum_{i'j'k'l'} \hat{D}(i i' j j' \, ;\, k k' l l')
  &= Tr(P^b_j P^a_i \rho_0 P^a_k P^b_{l} )\, \cr
  &= Tr( P^a_i \rho_0 P^a_k P^b_{j} )\, \delta_{jl}\, \cr
  &= Tr( \rho_0 P^a_i P^b_{j} )\, \delta_{ik}\delta_{jl}
   = p(a_i,b_j) \, \delta_{ik}\delta_{jl}
  \,,
 }
$$
using the cyclic property of the trace
and $P^b_l P^b_j = P^b_j \delta_{jl}$
in the middle line, and the
posited commutativity of $P^a_i$ and $P^b_j$ in the last.
\QED

Notice that $\hat{D}$ is only one of
many
decoherence
functionals which satisfy the desired conditions.
Instead of the product $P^a_i P^{a'}_{i'}$,
for example,
we could have any convex combination
of $P^a_i P^{a'}_{i'}$ and $P^{a'}_{i'} P^a_i$.
%% RDS: added following
(It seems unlikely that the most general $\hat{D}$ can be obtained in
 this manner, 
 though,
 because our ansatz here exhibits an extra decoherence not
 demanded by the physics; for example,
 $\hat{D}(ij;k'l):=\sum_{i'j'kl'}\hat{D}(ii'jj';kk'll')\propto\delta_{jl}$ 
 because
 $Tr(P^b_jP^a_i\rho_0 P^{a'}_{k'}P^b_l) \propto P^b_jP^b_l\propto\delta_{jl}$,
 even though $a\not=a'$.)

\REMARK
Even without commutativity, the above trace expressions would define
decoherence functionals $\hat{D}$ and $\hat{D}_{\alfa\beta}$ for
$\hat\Omega$ and the $\hat\Omega_{\alfa\beta}$, and the marginals of
$\hat{D}$ would still reproduce the $\hat{D}_{\alfa\beta}$.
In light of this, one might perceive the existence of a joint quantal
measure as reflecting most directly the existence for the
$\hat{D}_{\alfa\beta}$ of trace expressions involving operators in a
common Hilbert space 
(cf. [26]).
The commutativity 
would manifest itself, 
on this view, 
only in the fact that distinct $\alfa$-outcomes continue to decohere
independently of whether a $\beta$-measurement is made, and independently
of its outcome if it is.

% \REMARK
% Commutativity of $P^\alfa_i$ and $P^\beta_j$ is essential
% if one is to have decoherence on each $(\alfa, \beta)$ pair.

In the classical context, the existence of a joint probability
distribution $\hat\mu$ on $\hat{\Omega}$ is often described by saying
that one can find non-contextual hidden variables capable of 
reproducing the given system of experimental probabilities.
Adopting the same language, 
we can intepret 
Lemma \lsp in the following manner:
It is possible to attribute the correlations in the
EPRB setup to
non-contextual\footnote{$^\star$}
{By ``non-contextual'' we refer to the fact that the quantal measure
 $\hat\mu$ on $\hat{\Omega}$ is defined {\it independently of any
 measuring instruments or their settings}.  In this sense, one can say
 that a given measurement (if suitably designed) ``merely reveals'' a
 particular value of $\hat\mu$, without participating in its
 definition.
 (In saying this, we are not asserting that, in any individual instance,
 the measurement ``merely reveals'', for example, the location of the
 silver atom without affecting it.  
 This would be a much stronger claim, and possibly meaningless
 in a non-deterministic theory which
 provides no account of ``what would have happened'' in any individual
 instance, had the measurement not taken place.)}
hidden variables, 
so long as they are {\it quantal} hidden variables,
governed by
a decoherence functional
rather than a
classical probability distribution.\footnote{$^\dagger$}
{That a non-contextual quantal measure $\hat{\mu}$ exists where a 
 non-contextual classical measure cannot, implies that
 the corresponding decoherence functional $\hat{D}$ fails to be
 diagonal; for a decoherence functional is classical if it
 is diagonal.
 And indeed the $\hat{D}$ constructed above in the case of
 unitary quantum theory is not easily seen to possess off-diagonal
 matrix elements.  
 In the framework of the ``consistent histories'' point of view
 this means that the coarse-grained histories 
 specified by $(a_i,a'_{i'},b_j,b'_{j'})$ fail to decohere,
 and it is consequently not possible to assign
 probabilities simultaneously to all of the ``quantal hidden
 variables''.} 
%
%% Lemma \qjmso and \lsp,

\section{V. Weak positivity is not enough}
We have seen that the condition of strong positivity leads to the
Tsirel'son inequality.  Can the inequality be violated by a decoherence
functional that is only weakly positive (but otherwise observes the
conditions of theorem \thmB)?  That this is so can be seen simply by
noting the continuity of $Q$ in equation (13), and checking that
$\hat{D}_{sym}$ in equation (17) is not on the boundary of {\it
weak} positivity --- meaning that it assigns no set a measure of exactly
zero.  We have verified this for all of the $2^{16}-1$ non-empty subsets
of $\Omega$.

In fact, one might
go further and
ask whether the maximum 
possible value of $Q=4$ can be
attained by a weakly positive decoherence functional.  
The answer is yes, and there exist remarkably simple examples.
Here are the
elements of 
one such 
example
obtained by the {\tt lp\_solve}
linear programming solver [27].
$$
  \eqalign{
  \hat{D}(----;----) & = \hat{D}(+-+-;+-+-) = \hat{D}(+--+;+--+) \cr
  & = \hat{D}(-+-+;-+-+) = {1 \over 2} \cr
  -\hat{D}(---+;----) & = -\hat{D}(+-+-;+---) = \hat{D}(-+-+;-+--) \cr
  & = \hat{D}(---+;++--) = -\hat{D}(+--+;++--) \cr
  & = -\hat{D}(-+-+;++--) = \hat{D}(+--+;+-+-) \cr
  & = -\hat{D}(+--+;---+) = -\hat{D}(-+-+;---+) \cr
  & = \hat{D}(-+-+;+--+) = {1 \over 4}
  }
$$
The remaining elements which are not equal to one of the above by 
Hermiticity
are zero.
For this decoherence functional, one checks that
$$
   X(a,b) - X(a', b) + X(a, b') + X(a',b') = 4
$$

For consistency with theorem \thmB, strong positivity
must be violated by any $\hat{D}$ which violates the Tsirel'son bound.
One can check that the
above $\hat{D}$ does so, with four negative signs, four positive signs,
and eight zeros in its signature.

In the context of the Bell inequalities, then,
the strong positivity condition of quantum measure theory shows itself
to be 
much
stronger than the weak one.  To the extent that weak
positivity is physically acceptable, one can imagine a generalized form
of quantum mechanics (a generalized measure theory remaining at level
two)
which affords the maximum possible violation of the CHSHB inequality.
Strong positivity, in contrast, is as restrictive as ordinary quantum
mechanics in this respect.

One other feature of the above matrix $\hat{D}$ seems worthy of notice
here.  All the marginals of the form $\hat\mu(a_{\pm})$,
$\hat\mu(a'_{\pm})$, {\it etc.\ }take the value 1/2, which is recognizable
as
the only ``causal'' value.  That $\hat{D}$ yields $Q=4$ implies perfect
correlations (or anti-correlations) between $A$ and $B$, and any other
marginals than 1/2 would let Alya signal to Bai by manipulating the
settings of her analyzer.  But this could not happen with the above
$\hat{D}$ because ``non-signaling'' is built into the
%% decoherence
requirements we have imposed on it in equations (8),
which imply directly that $\sum_{j}p(a_i,b_j)=\sum_{j'}p(a_i,b'_{j'})$.

\section{VI. Conclusion}
One can view quantum mechanics as a dynamical schema that generalizes the
classical theory of stochastic processes in such a manner as to take into
account interference between pairs of alternatives.  
Within the framework appropriate to such a
view
--- that of ``quantal measure theory'' --- 
we have sought 
quantal 
analogs of some of
the
relationships that emerge in connection with correlated pairs of
spin-$\half$ particles when one
contemplates
tracing their behavior to the
dynamics of some underlying stochastic, but still classical,
variables (``hidden variables'').
One knows that
classically,
the existence of a joint probability measure on
the space of experimental outcomes is equivalent on one hand to the CHSHB
inequality, and on the other hand to screening off.
(This equivalence shows that the existence of hidden variables
is intimately linked to causality.)
Quantally, one might desire an analogous set of equivalences relating
(1) the
existence of a joint decoherence functional on the space
$\hat\Omega$
of experimental
outcomes; (2) Tsirel'son's inequality; and (3) some quantal causality
condition generalizing classical screening off.  We have shown --- assuming
strong positivity of the decoherence functional --- that (1) implies (2),
and
that (1) also implies (3) if the latter is represented by the candidate
condition (27).  A proof of the converse, that (3) implies (1), would
greatly strengthen the links with
%% quantum
causality.
We did not provide such
a proof
in general,
but we did show that (1) follows from
standard, unitary quantum mechanics with
spacelike commutativity.

It is perhaps worth emphasizing that, just as the CHSHB inequality follows
from the exceedingly general assumption of the existence of a joint
probability distribution
on $\hat\Omega$
(in effect, a probability distribution for non-contextual hidden variables),
making no statements concerning the nature of the classical dynamics
save that it is given by a probability measure on a suitable history-space,
so also
the Tsirel'son inequality
is a consequence
only of the bilinear
(level 2) structure of quantum theory.
We have
seen in fact
that it follows from the mere existence
of a
(strongly positive)
joint decoherence functional,
without making any assumption
that
the latter
has the form taken by ordinary unitary quantum mechanics.  The
inequality,
is
in this sense
a statement concerning the
predictive structure of quantum mechanics itself,
rather than anything to do with
any specific dynamical law.

If strong positivity is discarded, we can violate the Tsirel'son inequality
with a quantal ({\it i.e.\ }level two) measure, and
we have even seen
that the
``logical'' bound of 4 for the quantity $Q$
of equation (13)
can be achieved then.
The
corresponding non-local correlations are of interest in information theory,
since they would allow certain communication tasks to be performed with
fewer
classical bits transferred than are demanded in standard quantum mechanics
[28].  Quantal measure theory, or equivalently generalised
quantum mechanics, provides for such correlations, but only if strong
positivity is relaxed to weak positivity.  Whether this is
physically
appropriate
is doubtful,
however.
Apart from the Hilbert space constructions
that it affords, a compelling physical motivation for strong positivity
concerns
% has to do with
the composition of non-interacting sub-systems
[29]
% [R::Hartcnle:2004]
[15]:
Strong positivity is preserved under such composition
whereas
weak positivity is
not.\footnote{$^\flat$}
{This fact is closely related to the fact that tensor products of
 so called completely positive maps are also completely positive.}
(Might this difference lead to experimental tests that could distinguish
between the two types of positivity?)

For higher level measures, we speculate that imposing an analog of
strong positivity would lead to higher level inequalities
still weaker than (12),
but it is
beyond our current powers to pursue this idea since, beyond level two,
we lack the
analog
of the decoherence functional, in terms of which an
extension of the strong positivity condition could be framed.

Let us accept provisionally that the existence of a joint decoherence
functional is a necessary condition for relativistic causality.  Then we can
claim the following: If an EPRB-type experimental setup is ever found to
violate the Tsirel'son inequality, then all causal theories in the framework
of generalised quantum mechanics with a strongly positive decoherence
functional are contradicted.  However, as long as no superluminal signaling
is
seen, such an experimental result would not rule out causal generalised
quantum theories altogether,
if one were willing to
accept
that the world may be described by
decoherence functionals that are not strongly positive.
Another alternative would be to generalize to a higher order measure, in
which case the challenge would be to develop good dynamical models
within this at present loosely constrained class of theories.

A convincing quantal analog of the screening off condition would have an
interest going far beyond its relevance to experiments of the EPRB type.
In connection with quantum gravity, the condition of
``Bell causality'' was the guide that led to the family of
(classical) dynamical laws
derived in [30] for causal sets.
Screening off as such 
lacks a clear meaning 
against the backdrop of a dynamical causal structure,
but Bell causality
is perhaps as
close as one could have come to it in the causal set context.  For this
reason,
among others, it seems clear that
progress in identifying the correct quantal analog of classical
screening off would help point the way to a causality principle suitable for
the needs of quantum gravity.

The existence of a joint probability measure for
our 16-element sample space $\hat\Omega$ can be interpreted as the necessary
and sufficient condition for the existence of ``hidden variables'' which
determine
``non-contextually''
the measurement outcomes.  
That Bell's
inequality is violated in nature tells us that no such hidden variables are
possible {\it classically}.
Not so in quantal measure theory, 
however, 
and we described a
model in which the
``quantal hidden variables'' 
could be identified concretely with particle worldlines.  
Our main finished result in this paper was that the existence of such
variables can be seen as the reason for the Tsirel'son inequalities.
%
% although the
% significance of this for interpretations of quantum mechanics, if any,
% remains unclear.
% %% [[RDS  I've included a sentence on significance]]  
%
However, non-contextuality is only part of the story.
Whether such variables
can be ``causal'' as well as ``non-contextual'' is a question
whose
answer awaits a better understanding of the concept of ``quantal screening
off''.

\section{Acknowledgments}
We thank Rob Spekkens and Lucien Hardy for acquainting us with
reference [4],
Jonathan Barrett for reference [28]
and Tony Sudbery for reference [8].
We thank Tony Short for pointing out that the bound that $\hat{D}$
saturates in section V is not that of (13) but rather one of its
permutations.

D.C.\ thanks Hamilton College for its support.
J.H.\ was supported by Air Force grant AFOSR at UCSD.
S.M.\ was supported in part by an award from the Research Corporation.
D.R.\ was supported in part by the Marie Curie
  Research and Training Network ENRAGE (MRTN-CT-2004-005616).
The work of R.D.S.\ was partly supported
by NSF grant PHY-0404646 and
by a grant from the Office of Research and Computing of Syracuse University.

\ReferencesBegin

\ref[1]
J.S.~Bell, ``On the Einstein-Podolsky-Rosen paradox'',
\journaldata{Physics}{1}{195-200}{1964}.

\ref[2]
J.~Clauser, M.~Horne, A.~Shimony, and R.~Holt,
``Proposed experiment to test local hidden-variable theories'',
\journaldata{Phys.\ Rev.\ Lett.~}{23}{880-884}{1969}.

\ref[3]
J.S.~Bell,
 {\it Speakable and unspeakable in quantum mechanics:
   collected papers on quantum philosophy}
  (Cambridge University Press, 1987), chap.~4.

\ref[4]
Arthur Fine, ``Hidden Variables, Joint Probability, and the Bell
Inequalities'',
\journaldata{Phys.\ Rev.\ Lett.~}{48}{291-295}{1982}.

\ref[5]
Rafael D.~Sorkin,
``Quantum Mechanics as Quantum Measure Theory'',
   \journaldata{Mod.\ Phys.\ Letters~A}{9 {\rm (No.~33)}}{3119-3127}{1994},
\arxiv{gr-qc/9401003}.

\ref[6]
Lucien Hardy,
``Quantum mechanics, local realistic theories, and Lorentz-invariant
realistic theories'',
\journaldata{Phys.\ Rev.\ Lett.}{68}{2981-2984}{1992};
``Nonlocality for two particles without inequalities for almost all
entangled states'',
\journaldata{Phys.\ Rev.\ Lett.}{71}{1665-1668}{1993}.

\ref[7]
B.~Cirel'son, ``{Quantum generalisations of Bell's inequality}'',
\journaldata{Lett.\ Math.\ Phys.\ }{4}{93--100}{1980}.

\ref[8]
Paul Butterley, Anthony Sudbery, and Jason Szulc,
``Compatibility of subsystem states'',
\arxiv{quant-ph/0407227}.

\ref[9]
Joe Henson, ``Comparing causality principles'',
\linebreak
\arxiv{quant-ph/04010051}.

\ref[10] Chapter 7 of reference [3].

\ref[11]
The term ``stochastic Einstein locality'' 
was apparently coined in:
G. Hellman, ``Sto\-ch\-astic Einstein locality and the Bell theorems'',
\journaldata {Synthese} {53} {461-504} {1982}.
The formulations most suggesting equivalence to ``screening off'' can be
found elsewhere:
J.~Butterfield,
 ``{Outcome dependence and stochastic Einstein nonlocality}'',
 in
 {\it Logic and Philosophy of Science in Uppsala: Selected
  Papers from the 9th International Congress of Logic Methodology and
  Philosophy of Science} edited by
  D.~Prawitz and D.~Westerdahl
  (Kluwer, 1994),
   pp.~385-424.
%
% J.~Butterfield,  Outcome Dependence and Stochastic Einstein
% Nonlocality},  Logic and Philosophy of Science in Uppsala: Selected
% Papers from the 9th International Congress of Logic Methodology and
% Philosophy of Science} (D.~Prawitz and D.~Westerdahl, eds.),
% pp.~385-424, Kluwer, 1994.

\ref[12]
Rafael D.~Sorkin,
``Quantum Measure Theory and its Interpretation'', 
  in
   {\it Quantum Classical Correspondence:  Proceedings of the $4^{\rm th}$
    Drexel Symposium on Quantum Nonintegrability},
     held Philadelphia, September 8-11, 1994,
    edited by D.H.\ Feng and \hbox{B-L} Hu,
    pages 229--251
    (International Press, Cambridge Mass. 1997),
    \linebreak
   \arxiv{gr-qc/9507057}.

\ref[13]
Roberto B.~Salgado, ``Some Identities for the Quantum Measure and its
Generalizations'',
\journaldata{Mod. Phys. Lett.} {A17}{711-728}{2002},
\linebreak
\arxiv{gr-qc/9903015}.

\ref[14]
Xavier Martin, Denjoe O'Connor and Rafael D.~Sorkin,
``The Random Walk in Generalized Quantum Theory''
\journaldata {Phys. Rev.~D} {71} {024029} {2005}
\linebreak
\arxiv{gr-qc/0403085}.

\ref[15]
Fay Dowker, Raquel Garcia and Rafael D.~Sorkin,
``Hilbert space from quantum measure theory''
(in preparation).

\ref[16]
R.~B. Griffiths,
 ``Consistent histories and the interpretation of quantum mechanics'',
 \journaldata{J.~Statist.\ Phys.~}{36}{219-272}{1984}.

\ref[17]
R.~Omnes, ``Logical reformulation of quantum mechanics. 1. foundations'',
 \journaldata{J.\ Stat.\ Phys.} {53} { 893--932}{1988}.

\ref[18]
M.~Gell-Mann and J.~B. Hartle,
``Quantum mechanics in the light of quantum cosmology'',
in {\it Complexity, Entropy and the Physics of Information,
SFI Studies in the Sciences of Complexity, Vol VIII} (W.~Zurek, ed.),
pp.~150--173.
%%, Addison Wesley, Reading, 1990.

\ref[19]
J.B. Hartle, ``Spacetime Quantum Mechanics and the Quantum
 Mechanics of Spacetime'',
 in B. Julia and J. Zinn-Justin (editors),
 {\it Gravitation et Quantifications:
 Les Houches Summer School, session LVII, 1992}
 (Elsevier Science B.V. 1995),
\linebreak
\arxiv{gr-qc/9304006}.

% \ref[R::Isham:1994]
% C.~J.~Isham, ``Quantum logic and the histories approach to quantum theory''
% \journaldata{J.\ Math.\ Phys.~}{35}{2157}{1994},
%  \arxiv{gr-qc/9308006}.

\ref[20]
C.~J.~Isham, N.~Linden, and S.~Schreckenberg,
``The classification of decoherence functionals: an analogue of Gleason's
theorem''  \journaldata{J.\ Math.\ Phys.~}{35}{6360}{1994},
 \arxiv{gr-qc/9406015 }.

\ref[21]
D.~Craig, ``The geometry of consistency:
decohering histories in generalized quantum theory'',
\arxiv{gr-qc/9704031}.

\ref[22]
J.~D.~Maitland Wright,
``The structure of decoherence functionals for von Neumann quantum
histories''
 \journaldata{J.\ Math.\ Phys.~}{36}{5409-5413}{1995}.

% \ref[R::RW:1999]
% O.~Rudolph and J.~D.~M.~Wright,
% ``Homogeneous decoherence functionals in standard and history quantum mechanics''
%  \journaldata{Comm.\ Math.\ Phys.~}{204}{249}{1999},
%  \linebreak
%  \arxiv{math-ph/9807024}.

\ref[23]
Sukanya Sinha and Rafael~D.~Sorkin,
``A Sum-over-histories Account of an EPR(B) Experiment'',
   \journaldata{Found.\ of Phys.\ Lett.~}{4}{303-335}{1991}.

[24]
Fay Dowker and Joe Henson,
``A spontaneous collapse model on a lattice''
\journaldata{J.\ Stat.\ Phys.}{115}{1349}{2004},
 \arxiv{quant-ph/0209051}.

\ref[25]
Fay Dowker, Joe Henson and Rafael D.~Sorkin,
``Toward an Intrinsic Definition of Relativistic Causality''
(in preparation).

\ref[26]
O.~Rudolph and J.~D.~Maitland Wright,
``On tracial operator representations of quantum decoherence functionals''
 \journaldata{J.\ Math.\ Phys.~}{38}{5643-5652}{1997},
 \linebreak
 \arxiv{quant-ph/9706001}.

\ref[27]
{\tt http://groups.yahoo.com/group/lp\_solve}.

\ref[28]
J.~Barrett, N.~Linden, S.~Massar, S.~Pironio, S.~Popescu, and
%% \linebreak
D.~Roberts,
``Non-local correlations as an information theoretic resource'',
\linebreak
\arxiv{quant-ph/0404097};
D.~Rohrlich, S.~Popescu, ``Jamming non-local quantum correlations'',
\arxiv{quant-ph/9508001};
\linebreak
W.~Van Dam, PhD thesis, University of Oxford, Department of Physics,
\linebreak
{\tt http://web.mit.edu/vandam/www/publications.html}.

%% RDS: linebreaks modified to avoid overfull hboxes

\ref[29]
L.~Di\'{o}si, ``Anomalies of weakened decoherence criteria for quantum
histories'', \journaldata{Phys.\ Rev.\ Lett.}{92}{170401}{2004},
 \arxiv{quant-ph/0310181}.

% \ref[R::Hartle:2004]
% J.~B.~Hartle, ``Linear positivity and virtual probability'',
%  \arxiv{quant-ph/0401108}.

\ref [30]
David P.~Rideout and Rafael D.~Sorkin,
``A Classical Sequential Growth Dynamics for Causal Sets'',
\journaldata{Phys.\ Rev.\ D}{61}{024002}{2000},
\linebreak
\arxiv{gr-qc/9904062}.

\end

%: Outline mode stuff (put here so doesn't need to be commented out)

(prog1    'now-outlining
  (Outline
      "%------"
     "\f......"
      "%: ....."
   "\\\\message"
   "\\\\section"
   "\\\\appendi"
   "\\\\Referen"
   "\\\\Abstrac"
      "%:: ....."
   "\\\\subsectio"
   "\\\\ref......"))